\documentclass[preprint,12pt]{article}
\usepackage{amssymb}
\usepackage{amsmath}
\usepackage[numbers]{natbib}
\usepackage{graphicx}
\usepackage{geometry}
\usepackage{booktabs}
\usepackage{caption}
\usepackage{float}
\usepackage{url}
\usepackage{hyperref}
\usepackage{xcolor}
\usepackage{longtable}
\title{Multi-Scale Analysis of Nifty 50: Return Characteristics, Valuation Dynamics and  Market Complexity (1990–2024)}

\author{Chandradew Sharma \\
  Department of Physics, Birla Institute of Technology and Science, \\
  Pilani K K Birla Goa Campus, NH-17 B, Zuarinagar, Goa 403726, India}

\begin{document}
\maketitle

\begin{abstract}

This study presents a unified, distribution-aware, and complexity-informed framework for understanding equity return dynamics in the Indian market, using 34 years (1990--2024) of Nifty 50 index data. Addressing a key gap in the literature, we demonstrate that the price-to-earnings (P/E) ratio, as a valuation metric, may probabilistically map return distributions across investment horizons spanning from days to decades.

Return profiles exhibit strong asymmetry. One-year returns show a 74\% probability of gain, with a modal return of 10.67\% and a reward-to-risk ratio exceeding 5. Over long horizons, modal CAGRs surpass 13\%, while worst-case returns remain negative for up to ten years—defining a historical ``trapping period.''  This horizon shortens to six years in the post-1999 period, reflecting growing market resilience.

Conditional analysis of the P/E ratio reveals regime-dependent outcomes. Low valuations (P/E $<$ 13) historically show zero probability of loss across all horizons, while high valuations (P/E $>$ 27) correspond to unstable returns and extended breakeven periods.

To uncover deeper structure, we apply tools from complexity science. Entropy, Hurst exponents, and Lyapunov indicators reveal weak persistence, long memory, and low-dimensional chaos. Information-theoretic metrics---including mutual information and transfer entropy---confirm a directional and predictive influence of valuation on future returns.

These findings offer actionable insights for asset allocation, downside risk management, and long-term investment strategy in emerging markets. Our framework bridges valuation, conditional distributions, and nonlinear dynamics in a rigorous and practically relevant manner.

\end{abstract}



\section{Introduction}  
\label{sec:introduction}

Understanding equity market returns remains a central and persistent challenge in financial economics. Traditional models often overlook nonlinear dynamics, valuation-dependent behavior, and risks unfolding over long horizons, especially in emerging markets. This study uniquely addresses these gaps by developing an integrated, high-resolution, multi-scale framework that simultaneously incorporates return distributions, valuation regimes, and complexity measures, applied to the Indian equity market through the Nifty~50 index from 1990 to 2024.

Unlike prior works that treat valuation, returns, and complexity in isolation, our approach unifies these perspectives by blending insights from econophysics, financial econometrics, and nonlinear time series analysis~\cite{mandelbrot1963variation, Fama1965, liu1999statistical, banerjee2014return, Zhou2014, zhou2016distribution, Duan2018, Nadarajah2023, Dutta2023, Haddari2024}. This integration enables a richer understanding of how valuation extremes, earnings momentum, and market complexity jointly influence return asymmetry, trapping horizons, and investment performance. Such insights hold direct relevance for portfolio managers and risk professionals navigating the evolving dynamics of emerging equity markets.

Building on this interdisciplinary foundation, the nonlinear, multifractal, and chaotic methods applied here are closely aligned with analytical techniques in fluid mechanics and complex dynamical systems, underscoring the interdisciplinary nature of market behavior and the applicability of tools from turbulence and nonlinear dynamics to financial markets.

Leveraging these advanced analytical tools, we focus on the post-reform period (1999--2024), we empirically analyze the Nifty~50 index and its price-to-earnings (P/E) ratio (Figure~\ref{fig:Nifty50ts}), illustrating how valuation spikes have preceded major downturns such as the Dot-Com bust, the 2008 financial crisis, and the COVID-19 pandemic drawdown. The exceptional valuation surge in 2020--2021—driven primarily by liquidity and capital inflows despite subdued earnings—further highlights the nonlinear and regime-dependent nature of modern markets.

To complement valuation analysis, we introduce a novel daily proxy for earnings per share (EPS) defined as:  
\[
    \text{EPS}_t = \frac{P_t}{(P/E)_t},
\]  
and examine its trailing one-year growth. The resulting EPS growth distribution (Figure~\ref{fig:Nifty501eps}) is notably skewed and heavy-tailed, reflecting burst-like earnings momentum during market recoveries and underscoring the inadequacy of Gaussian assumptions~\cite{Singh2023, Joshi2023, JoshiKulkarni2023, Patel2024}.

While valuation extremes provide valuable signals, they alone do not fully explain market drawdowns or recoveries. Instead, it is the combined behavior of valuation and earnings momentum that better accounts for market resilience or vulnerability. For example, the high post-COVID P/E ratios were supported by strong earnings rebounds, enabling sustained bullish phases despite macroeconomic uncertainty (Figure~\ref{fig:Nifty50tscrash}).

Next, we characterize return distributions across multiple horizons, from daily to over a decade. One-day returns (1990--2024) exhibit mild positive skewness, a reward-risk ratio of 1.26, and 56\% positive days (Figure~\ref{fig:Nifty50onedayreturn}). By contrast, one-year returns show broader asymmetry, with a 74\% probability of gains, a mode of 10.67\%, and a reward-risk ratio exceeding 5.0 (Figure~\ref{fig:Nifty501yreturn}), emphasizing the benefits of long-term investment patience.

Extending this to horizons up to twelve years (Figure~\ref{fig:Nifty50mmmreturn90}), we identify a ``trapping horizon'' of roughly ten years during which worst-case returns remain negative, delaying compounding benefits. Multi-year compound annual growth rate (CAGR) summaries (Table~\ref{tab:Nifty50cagr}) quantify this phenomenon, with modal CAGRs exceeding 10\% and worst-case scenarios turning positive only after ten years. Comparing the full period with the post-1999 sub-period reveals a shortened trapping horizon of six years and faster recovery cycles, suggesting increased market maturity and resilience.

We then investigate the complexity underlying market dynamics using entropy, chaos, and fractal measures (Section~\ref{sec:pmf_complexity}). Results reveal multifractality, weak persistence, and low-dimensional chaos, highlighting nonlinear structures beyond classical models.

\begin{figure}[t]
\centering
\includegraphics[width=1.0\linewidth, keepaspectratio]{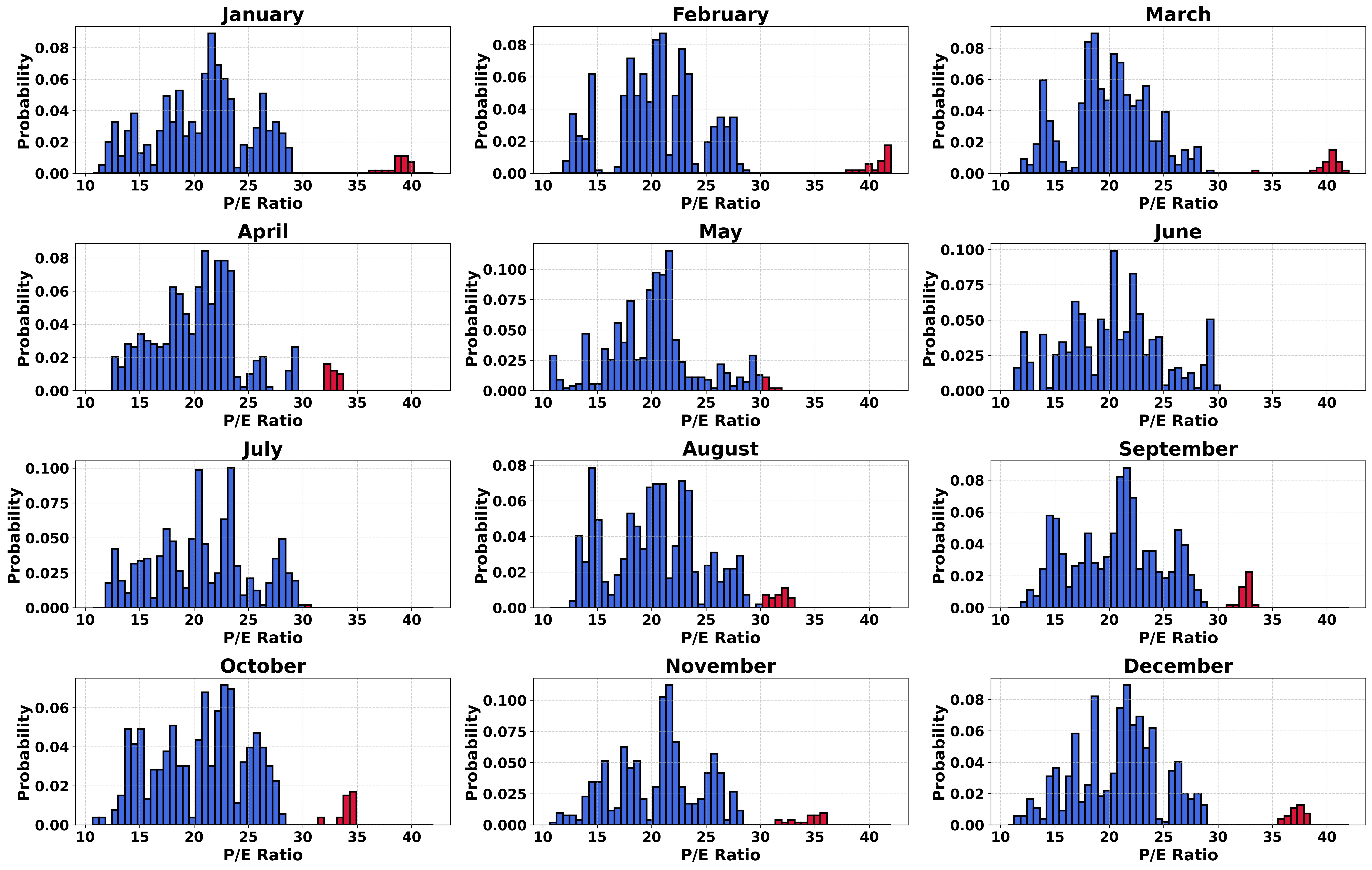}
\caption{Monthly PMF of Nifty 50 P/E Ratio}
\label{fig:Nifty50PEm}
\end{figure}

To capture valuation dynamics more deeply, we estimate empirical daily P/E ratio distributions (1999--2024) via Freedman--Diaconis binning (Figure~\ref{fig:Nifty50PE}). The sharply peaked distribution centers at 21.02, with well-defined \(\pm 1\sigma\) and \(\pm 2\sigma\) bands that serve as practical guides for regime-aware portfolio decisions. Monthly P/E distributions (Subsection~\ref{sec:PEdistm}) reveal seasonal asymmetries and multimodalities linked to macro-financial cycles (Figure~\ref{fig:Nifty50PEm}).

Building upon this distributional analysis, we further investigate the underlying complexity and nonlinear characteristics of the P/E ratio using a suite of entropy measures—including Shannon, Tsallis, Sample, and Permutation entropy—demonstrating a mixture of structure and randomness in valuation signals (Table~\ref{tab:entropy_results}). Generalized Hurst exponents (Section~\ref{sec:hurstexp}, Table~\ref{tab:GH}) confirm scale-dependent persistence and multifractality, while a five-dimensional Lyapunov spectrum (Subsection~\ref{sec:lyapunovexp}) reveals chaotic yet low-dimensional dynamics with a Kolmogorov--Sinai entropy rate of approximately 0.41 and attractor dimension near 4.07 (Table~\ref{tab:lyap_spectrum}).

Extending beyond intrinsic valuation properties, we analyze the information-theoretic analysis of valuation and returns (Section~\ref{sec:mutualinfo}) uncovers persistent nonlinear dependence across multiple lags. Normalized Mutual Information (NMI) remains elevated above 0.4 for up to 15 lags and stabilizes near 0.28 by lag 50 (Figure~\ref{fig:nmi_plot}), motivating causal investigations.

Following this evidence, we apply  Mutual Information and Transfer Entropy (Subsection~\ref{sec:MI_TE}), we confirm a modest but significant directional influence from P/E ratios to future returns, surpassing symmetric association measures~\cite{DimpflPeter2013, DimpflPeter2014, Harre2014, KorbelJiangZheng2019, YaoLi2020}.

Finally, conditional return distributions across valuation bands (Subsection~\ref{sec:cond_return_givenPE}) reveal regime-dependent asymmetries in return probabilities over horizons from one to seven years~\cite{Ghysels2020, NoretsPelenis2022, MaynardShimotsuKuriyama2023, marsili1999scaling, Yan2022_PEN}. Reward-Risk Ratio (RRR) analysis highlights a transition from near ``no risk'' regimes at low valuations to significantly adverse regimes at high valuations.

Synthesizing these insights, we propose a valuation-conditioned, distribution-based framework for portfolio construction and risk management that aligns allocation with prevailing valuation conditions, return asymmetries, and long-horizon dynamics (Sections~\ref{sec:RRR} and \ref{sec:summary}). This approach offers a rigorous, complexity-aware alternative to traditional mean-variance models, with practical implications for asset allocators and risk managers in emerging markets like India.

The structure of the paper is as follows: Section ~\ref{sec:Rnifty50} provides a comprehensive empirical analysis of Nifty 50 returns, valuation regimes, and earnings momentum, establishing the foundational basis for the study’s subsequent analyses. Section~\ref{sec:complexityPE} details the nonlinear complexity analysis of the P/E ratio, including entropy measures, fractal properties, and information-theoretic dependencies with returns. Section~\ref{sec:returngivenPEratio} presents an analysis of conditional return distributions across valuation regimes, highlighting the valuation-dependent asymmetries and associated reward-risk dynamics. Section~\ref{sec:result_discussion} presents a unified empirical and dynamical analysis of valuation regimes, return asymmetries, and market complexity in the Nifty~50 index from 1990 to 2024. Finally, Section~\ref{sec:conclusion}  concludes by summarizing the key contributions and outlining directions for future research.

In summary, this manuscript introduces a novel, unified framework that combines return distributions, nonlinear complexity, and valuation signals. Our findings provide both theoretical insights and actionable guidance for long-horizon investors navigating the intricate behavior of the Indian equity market.

\section{Understanding Historical and Prospective Returns in the Nifty 50 Index}
\label{sec:Rnifty50}

\begin{figure}[H]
\centering
\includegraphics[width=1.0\linewidth, keepaspectratio]{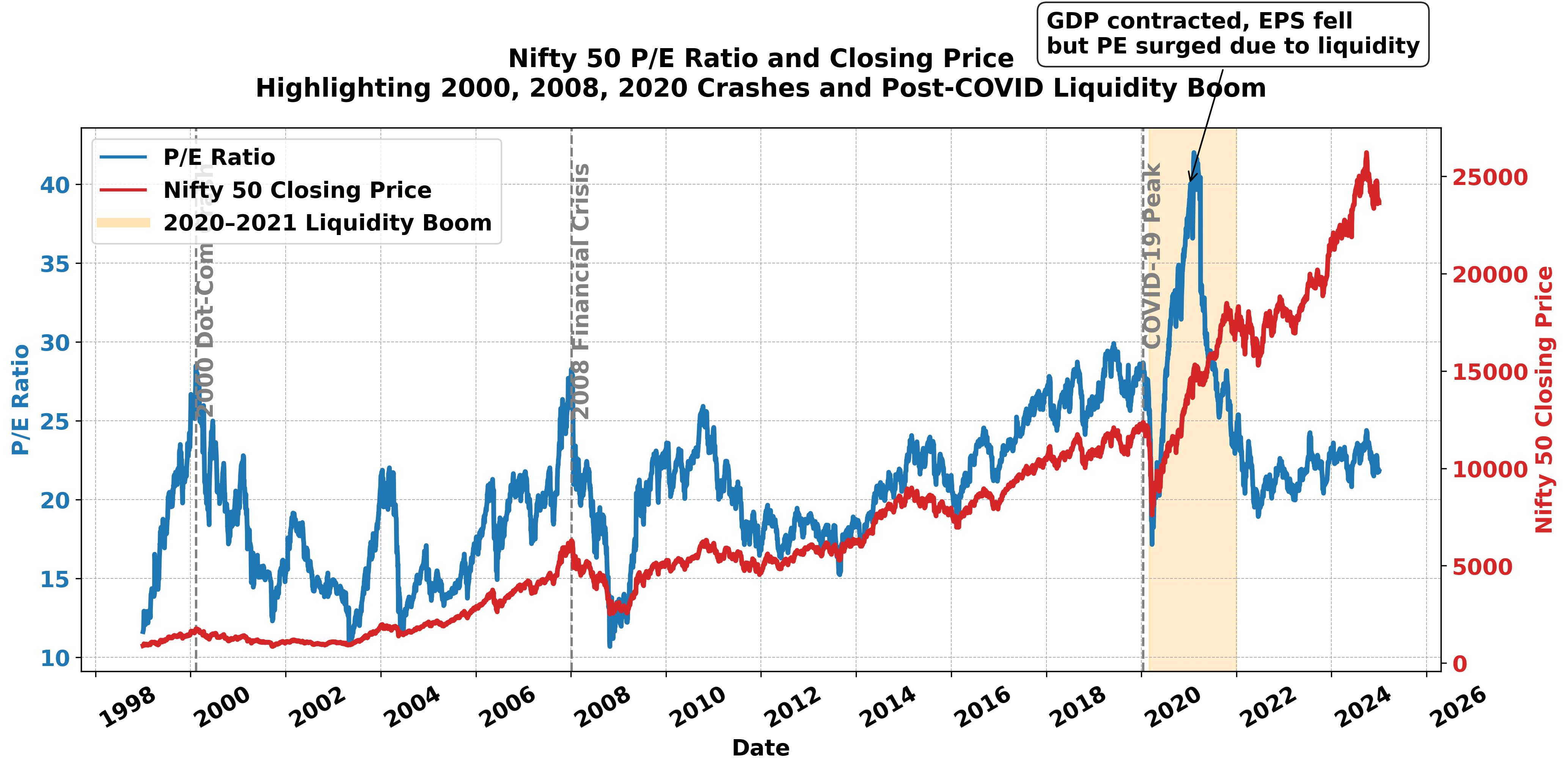}
\caption{Nifty 50 index closing price and P/E ratio (1999–2024).}
\label{fig:Nifty50ts}
\end{figure}

Forecasting equity returns across investment horizons—from days to decades—remains a central and challenging problem in financial economics. While valuation metrics such as the price-to-earnings (P/E) ratio are widely employed, their predictive power is often questioned, particularly at short horizons. Moreover, the presence of structural regime shifts, behavioral biases, and policy interventions complicates their long-term reliability. This study addresses a critical gap by examining whether a single valuation metric, embedded within a probabilistic and multi-scale framework, can meaningfully capture the distribution of future returns and the complexity of underlying earnings dynamics.

To explore this question empirically, we focus on the Nifty 50 index, a key benchmark representing the Indian equity market, and its associated exchange-traded funds (ETFs).   Understanding returns in this context requires a detailed examination of the historical interplay between market valuations and fundamental economic variables. Accordingly, we analyze daily closing prices of the Nifty 50, sourced from the National Stock Exchange of India (NSE)~\cite{NSE2024}, spanning July 3, 1990, through December 31, 2024. Corresponding P/E ratio data are available from January 1, 1999, onward. Figure~\ref{fig:Nifty50ts} depicts the evolution of the Nifty 50 index and its P/E ratio over this period. Notably, the P/E ratio consistently signals valuation extremes, with pronounced peaks preceding major market downturns such as the Dot-Com bust in 2000, the 2008 Financial Crisis, and the COVID-19 pandemic induced shock in early 2020. 

However, the post-COVID period represents a marked departure from these historical patterns. Despite a sharp contraction in GDP growth, subdued corporate earnings, and heightened macroeconomic uncertainty, the Nifty 50’s P/E ratio surged to unprecedented levels between mid-2020 and 2021, reflecting a valuation expansion that appeared decoupled from underlying earnings fundamentals.

This seemingly paradoxical behavior can be attributed to a confluence of domestic and global factors. In response to the pandemic-induced economic slowdown, the Reserve Bank of India implemented aggressive monetary easing, including a 75 basis point reduction in the repo rate, a cut in the cash reserve ratio, and substantial liquidity injections via open market operations and the Standing Deposit Facility. Simultaneously, fiscal stimulus measures bolstered aggregate demand~\cite{RBI2020}. Concurrently, strong foreign institutional investor (FII) inflows, attracted by India’s relative macroeconomic resilience and a global search for yield, fueled a sharp equity rally. Consequently, asset prices rose rapidly despite lagging corporate earnings, highlighting the limitations of relying solely on the P/E ratio as a forward-looking valuation measure.

\subsection{Earnings Proxy and Growth Distribution}
\label{subsec:epsgrowth}

To further investigate this divergence between market valuations and fundamentals, we construct a daily proxy for earnings performance of the Nifty 50 index by leveraging the relationship between the closing price and the P/E ratio. Specifically, earnings per share (EPS) on trading day \( t \) is estimated as:
\[
\text{EPS}_t = \frac{P_t}{(P/E)_t},
\]
where \( P_t \) denotes the closing price and \( (P/E)_t \) the corresponding price-to-earnings ratio. This transformation yields a high-frequency series of implied EPS values that reflect evolving market sentiment and earnings expectations.

To capture momentum in corporate profitability, we compute the trailing one-year EPS growth rate as the percentage change in EPS relative to its value 252 trading days prior:
\[
\text{Trailing 1-Year EPS Growth} (\%) = \left( \frac{\text{EPS}_t - \text{EPS}_{t-252}}{\text{EPS}_{t-252}} \right) \times 100.
\]
This rolling annual growth metric provides a timely indicator of shifts in earnings dynamics and complements valuation ratios in assessing broader market conditions.

\begin{figure}[H]
\centering
\includegraphics[width=1.0\textwidth]{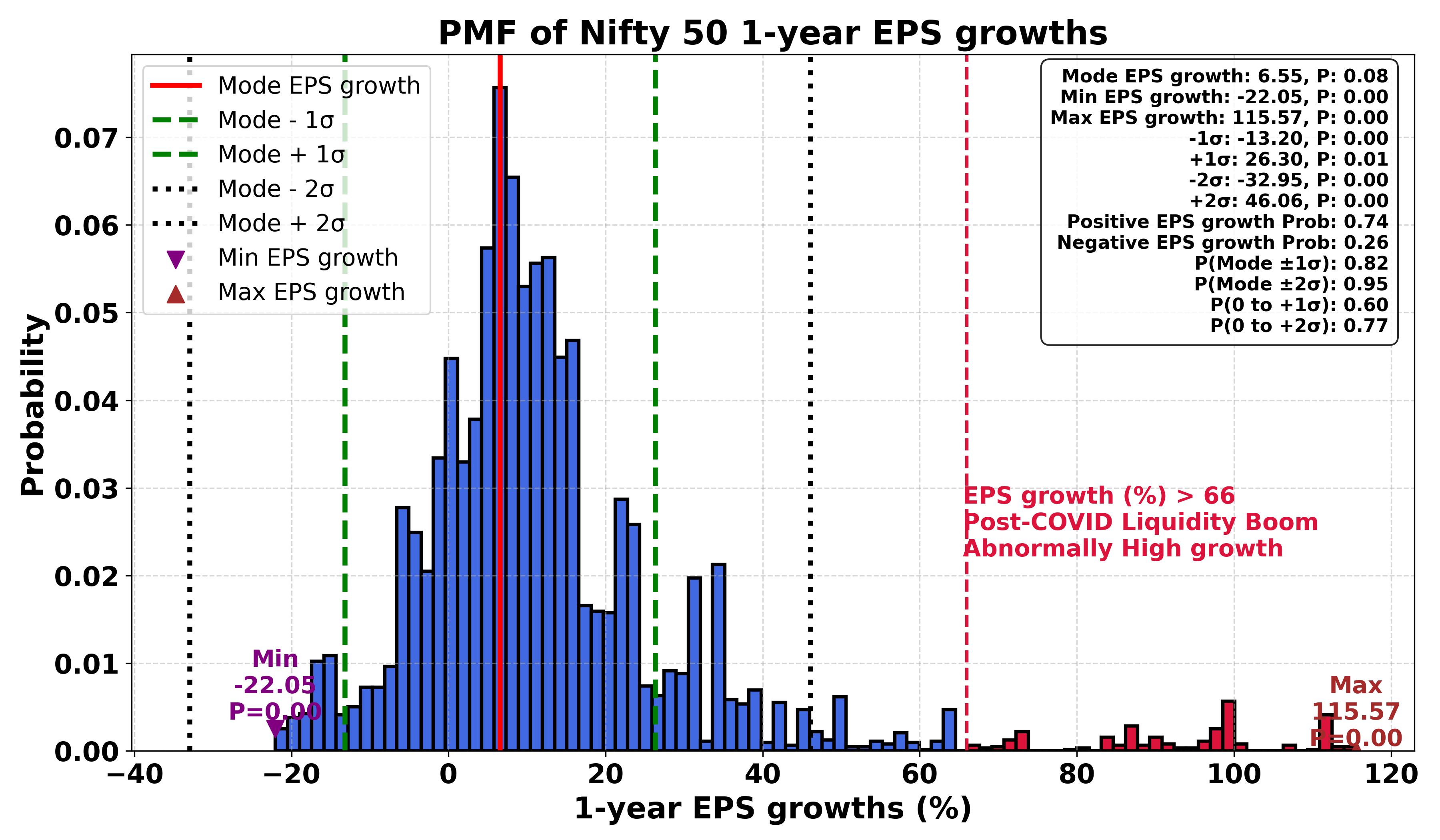}
\caption{Probability mass function (PMF) of trailing 1-year EPS growth for the Nifty 50 index (2000–2024).}
\label{fig:Nifty501eps}
\end{figure}

Figure~\ref{fig:Nifty501eps} presents the empirical PMF of trailing one-year EPS growth from 2000 to 2024. The distribution exhibits a strong central tendency around 6.55\%, but also displays significant positive skewness and a pronounced right tail, indicative of fat-tailed behavior. This asymmetry reflects rare but substantial earnings surges, particularly during the post-COVID recovery period, driven by reopening effects, pent-up demand, and accelerated digital transformation across industries. Some firms recorded EPS growth exceeding 65\%, underscoring the heterogeneous and asymmetric nature of corporate earnings outcomes. Such distributional features challenge the common Gaussian assumptions often employed in financial modeling and emphasize the necessity of incorporating nonlinear and probabilistic frameworks when evaluating earnings risk, valuation, and return potential~\cite{Singh2023, Joshi2023, JoshiKulkarni2023, Patel2024}.

\subsection{Joint Dynamics of Price, Valuation, and Earnings}
\label{subsec:jointdynamics}

\begin{figure}[H]
\centering
\includegraphics[width=1.0\textwidth]{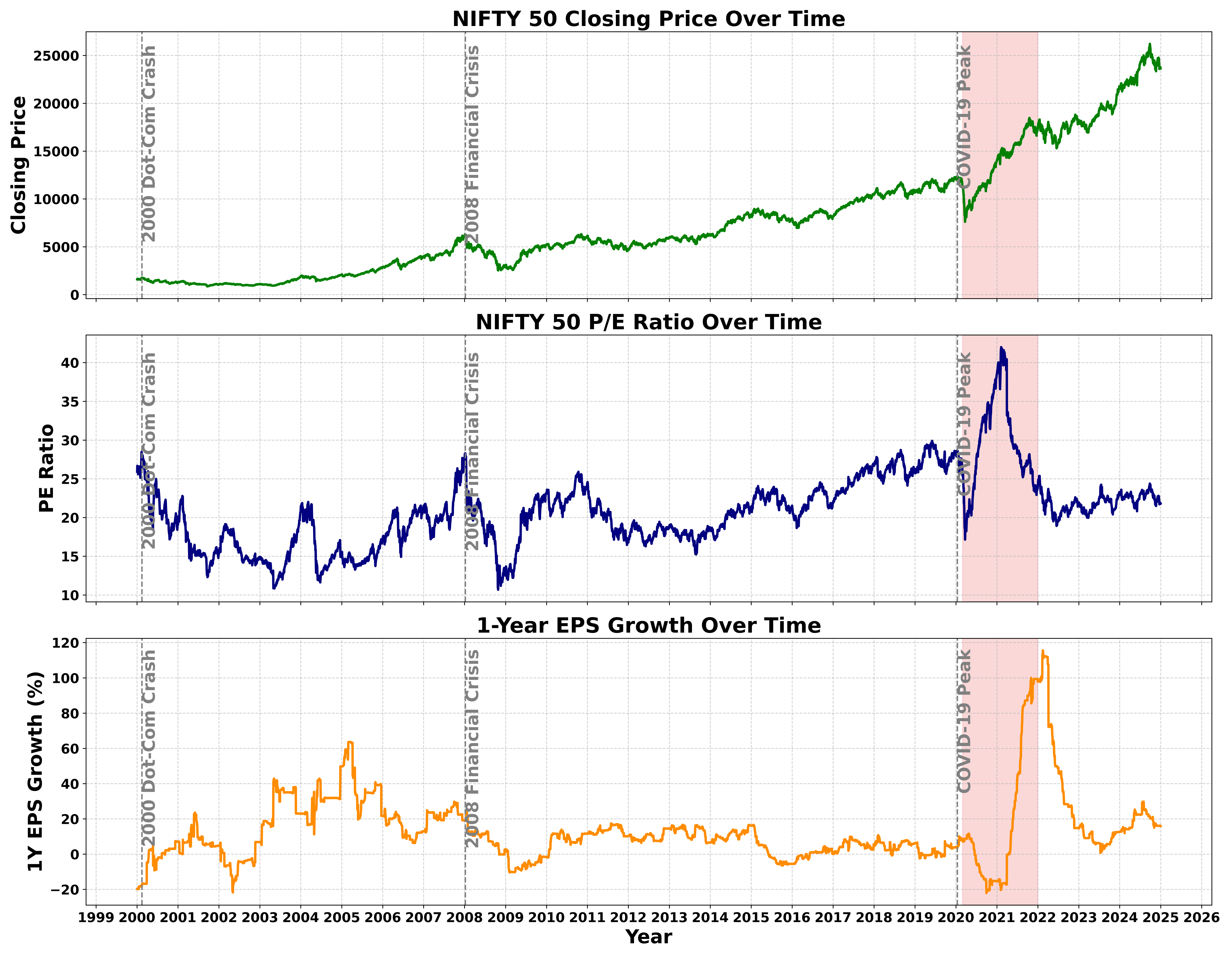}
\caption{Nifty 50 index closing price, P/E ratio, and trailing 1-year EPS growth (2000–2024).}
\label{fig:Nifty50tscrash}
\end{figure}

Figure~\ref{fig:Nifty50tscrash} presents the joint time-series evolution of the Nifty 50 index closing price, P/E ratio, and trailing one-year EPS growth from 2000 to 2024, offering critical insights into the interaction between valuation levels, earnings momentum, and price dynamics. Periods of elevated valuations frequently coincided with either decelerating or accelerating earnings growth, shaping the market’s risk-return profile across horizons. For instance, the Dot-Com bust and the Global Financial Crisis were preceded by high P/E ratios alongside weakening earnings, while the post-COVID rally was marked by historically elevated valuations sustained by a sharp rebound in EPS growth. Notably, the divergence between prices and earnings during liquidity-driven episodes—such as the 2020–2021 phase—highlights the role of monetary and fiscal interventions, investor sentiment, and global capital flows in amplifying market movements. These joint dynamics underscore that while high valuations may signal correction risks, the trajectory and resilience of earnings remain central to understanding return distributions and systemic vulnerability over multiple time scales.

\subsection{One-Day Return Distribution}
\label{subsec:onedayreturns}

\begin{figure}[H]
    \centering
    \includegraphics[width=1.0\textwidth]{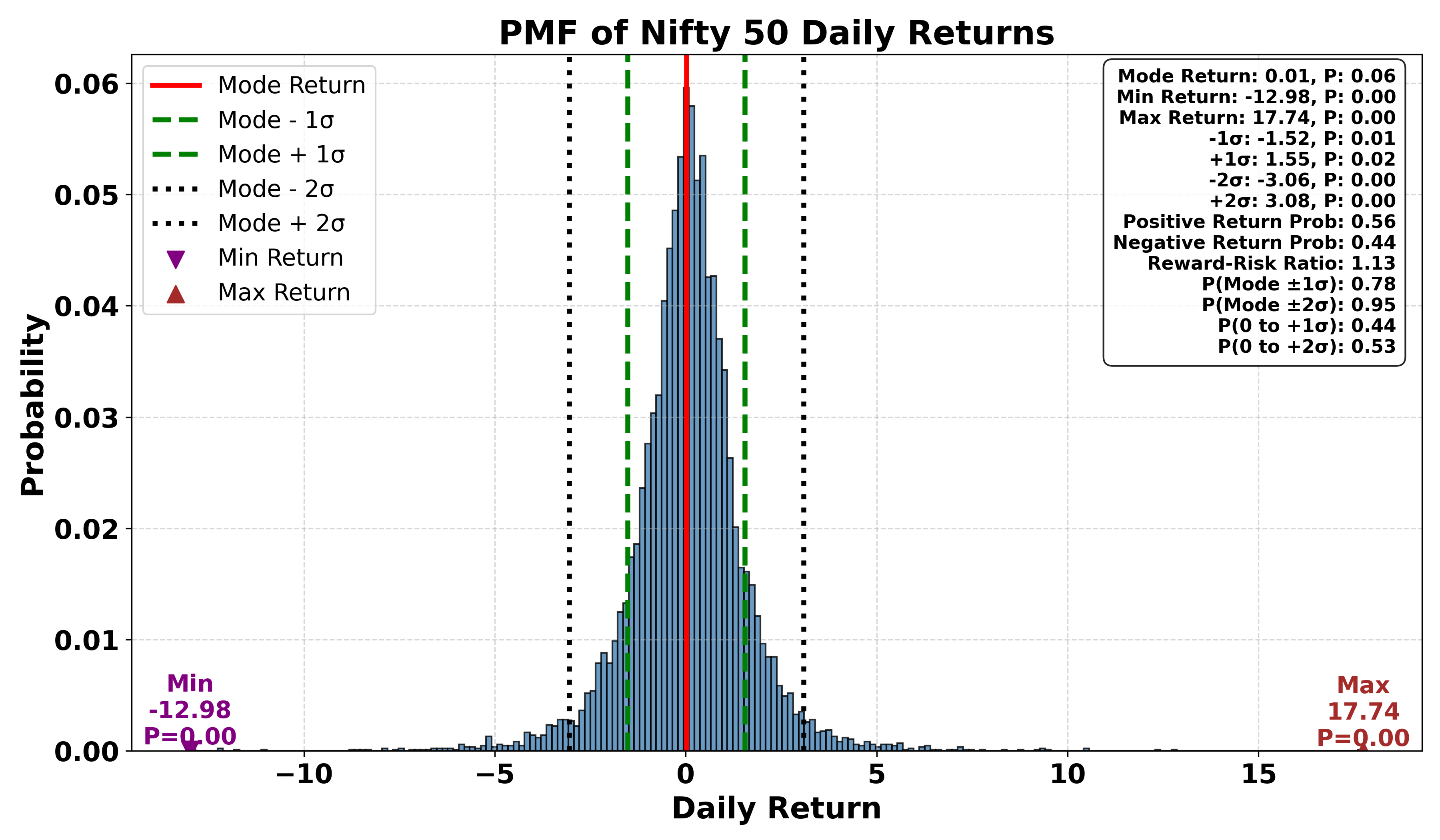}
    \caption{PMF of Nifty 50 One-Day Returns (1990--2024).}
    \label{fig:Nifty50onedayreturn}
\end{figure}

To develop a more comprehensive understanding of return dynamics in Indian equity markets, this study adopts a multi-horizon, distributional approach~\cite{mandelbrot1963variation,  Fama1965,  liu1999statistical,  banerjee2014return,   Zhou2014,   zhou2016distribution,  Duan2018, Nadarajah2023, Dutta2023, Haddari2024} grounded in historical Nifty 50 data from January 1, 1990, to December 31. For robust and adaptive estimation of the PMF structure, we employ the Freedman–Diaconis rule~\cite{scott1979, freedman1981, rudemo1982, birge2006} to determine bin widths. This method ensures that the binning is sensitive to both sample size and variability, thus preserving important distributional features without over-smoothing. By analyzing returns across varying investment horizons—from one day to one year—we move beyond static valuation metrics to assess the probabilistic nature of returns, the asymmetry of outcomes, and evolving reward-risk profiles over time. 

At the shortest horizon, we examine the empirical distribution of one-day returns, which reflect the market’s immediate response to overnight developments, macroeconomic news, and intraday trading dynamics. Formally, the \(i\)-day return \( R_{t+i} \) is defined as the percentage change from the closing price on day \( t \) to that on day \( t+i \):
\[
R_{t+i} = \frac{P_{t+i} - P_t}{P_t} \times 100,
\]
where \( P_t \) and \( P_{t+i} \) denote the closing prices on trading days \( t \) and \( t+i \), respectively, with \( i \in \{1, 5, 10, 21, 252\} \) representing investment horizons of one day, one week, two weeks, one month, and one year, respectively.

To evaluate the asymmetry of return outcomes, we define the \textit{expected positive return} \( \mathbb{E}[R_{+}] \) as the weighted average of all strictly positive returns in the empirical PMF, and the \textit{expected negative return} \( \mathbb{E}[R_{-}] \) as the weighted average (in absolute value) of all strictly negative returns:
\begin{equation}
\mathbb{E}[R_{+}] = \sum_{x_j > 0} x_j \cdot p_j,
\label{eq:expected_positive_return}
\end{equation}
\begin{equation}
\mathbb{E}[R_{-}] = \sum_{x_j < 0} |x_j| \cdot p_j,
\label{eq:expected_negative_return}
\end{equation}
where \( x_j \) represents return values and \( p_j \) their empirical probabilities. The ratio of these two expectations yields the PMF-based \textit{Reward-Risk Ratio}~\cite{RewardRiskTurkish2023}:
\begin{equation}
\text{Reward-Risk Ratio} = \frac{\mathbb{E}[R_{+}]}{\mathbb{E}[R_{-}]},
\label{eq:reward_risk_ratio}
\end{equation}
providing a probabilistically grounded measure of directional return asymmetry that complements traditional risk measures based on standard deviation.

Figure~\ref{fig:Nifty50onedayreturn} presents the empirical PMF of one-day Nifty 50 returns. The distribution is sharply peaked near zero, with a modal (most probable) return of approximately \textbf{0.01\%} occurring with a probability of \textbf{6\%}, reflecting the frequency of minimal daily movements. The PMF exhibits a modest positive skew, with \textbf{56\%} of daily returns being positive and a reward-risk ratio of approximately \textbf{1.26}, suggesting a slight upward bias. Most of the upside is driven by frequent moderate gains: \textbf{44\%} of returns fall between \(0\%\) and \(+1\sigma\), and \textbf{53\%} between \(0\%\) and \(+2\sigma\). Dispersion is captured by standard deviation bands, with \(\pm1\sigma\) spanning \(-1.52\%\) to \(+1.55\%\) (capturing \textbf{78\%} of observations) and \(\pm2\sigma\) covering \(-3.06\%\) to \(+3.08\%\) (capturing \textbf{95\%} of returns). Tail events are rare but significant, with extremes ranging from \textbf{-12.98\%} to \textbf{+17.74\%}, often associated with global crises or sudden policy shifts.

\begin{figure}[H]
    \centering
    \includegraphics[width=1.0\textwidth]{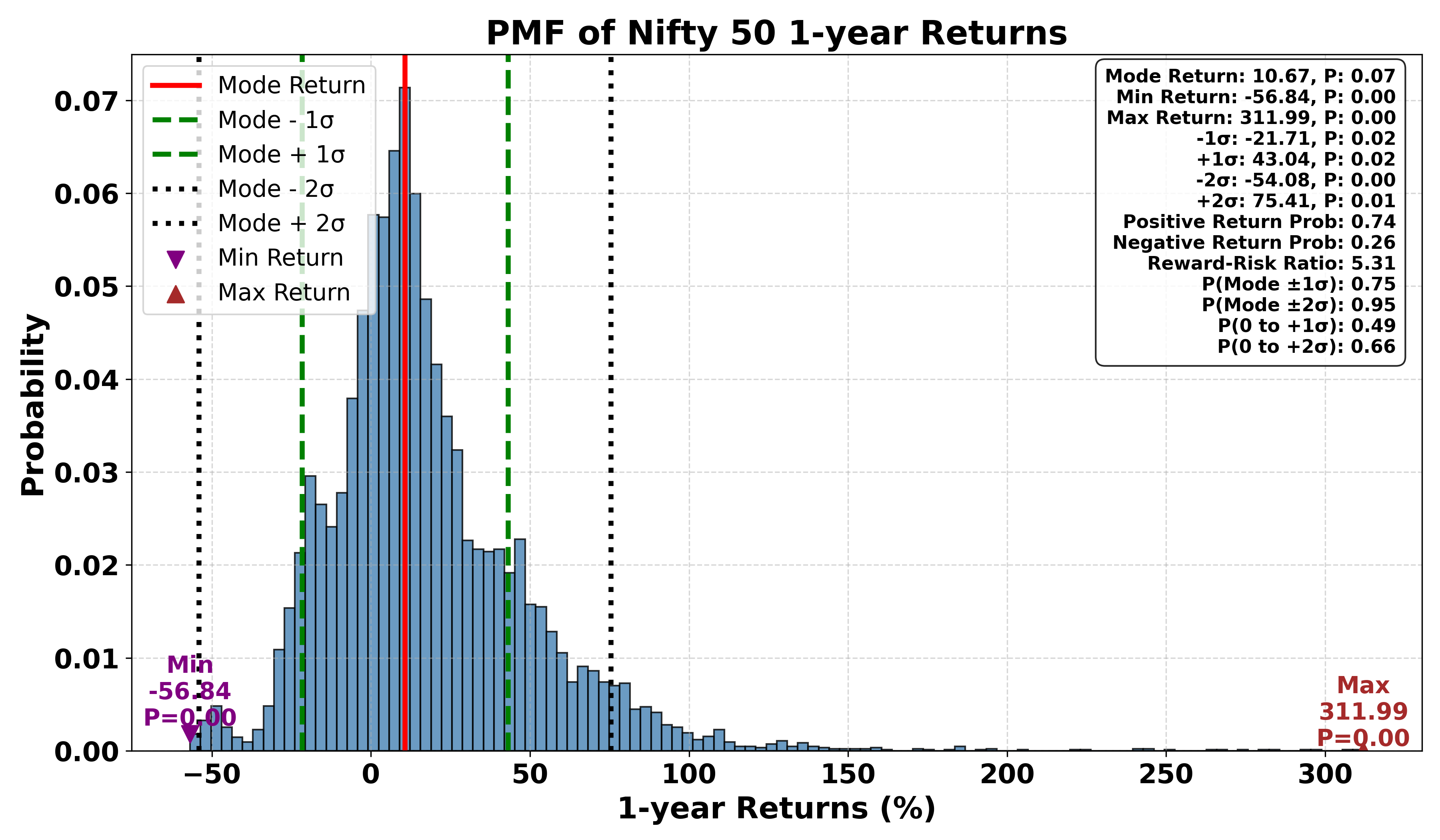}
    \caption{PMF of Nifty 50 One-Year Returns (1990--2024).}
    \label{fig:Nifty501yreturn}
\end{figure}

Extending the horizon, Figure~\ref{fig:Nifty501yreturn} illustrates the PMF of one-year forward returns. The modal one-year return is \textbf{10.67\%}, occurring with a \textbf{7\%} probability, indicating a historical tendency for moderate annual gains. The distribution is broader and more positively skewed than its short-term counterpart, with outcomes ranging from a minimum of \textbf{-56.84\%} to a maximum of \textbf{311.99\%}. The \(\pm1\sigma\) interval spans from \textbf{-21.71\%} to \textbf{+43.04\%}, covering \textbf{75\%} of returns, while the \(\pm2\sigma\) band from \textbf{-54.08\%} to \textbf{+75.41\%} captures \textbf{95\%}. The probability of a positive one-year return is \textbf{74\%}, and the reward-risk ratio rises to \textbf{5.31}, underscoring the long-term bullish bias of Indian equities. Similar to the short-term horizon, moderate gains dominate the upside: \textbf{49\%} of returns fall between \(0\%\) and \(+1\sigma\), increasing to \textbf{66\%} within \(0\%\) to \(+2\sigma\).

\subsection{One-Year Return Distribution}
\label{subsec:oneyearreturns}

\subsection{Multi-Horizon Return Distributions}
\label{subsec:multihorizon}

To complement the distributional analysis of one-year returns and provide a broader, time-consistent perspective on return dynamics, we extend our investigation across multiple investment horizons. Specifically, we evaluate  returns of the Nifty 50 index over holding periods of 1-day, 1-week, 2-week, 1-month, 3-month, 6-month, 1-year, and annually from 2 to 12 years, using daily closing price data spanning January 1, 1990, to December 31, 2024. This expanded analysis captures the evolving behavior of return distributions over time, offering insights into how dispersion, skewness, and modal outcomes shift across short-, medium-, and long-term horizons. The full set of empirical PMFs for each horizon is presented in the Appendix ( Figure~\ref{fig:Nifty50as90} ), enabling a granular view of probabilistic return profiles across temporal scales.

\begin{figure}[H]
     \centering
     \includegraphics[width=1.0\textwidth]{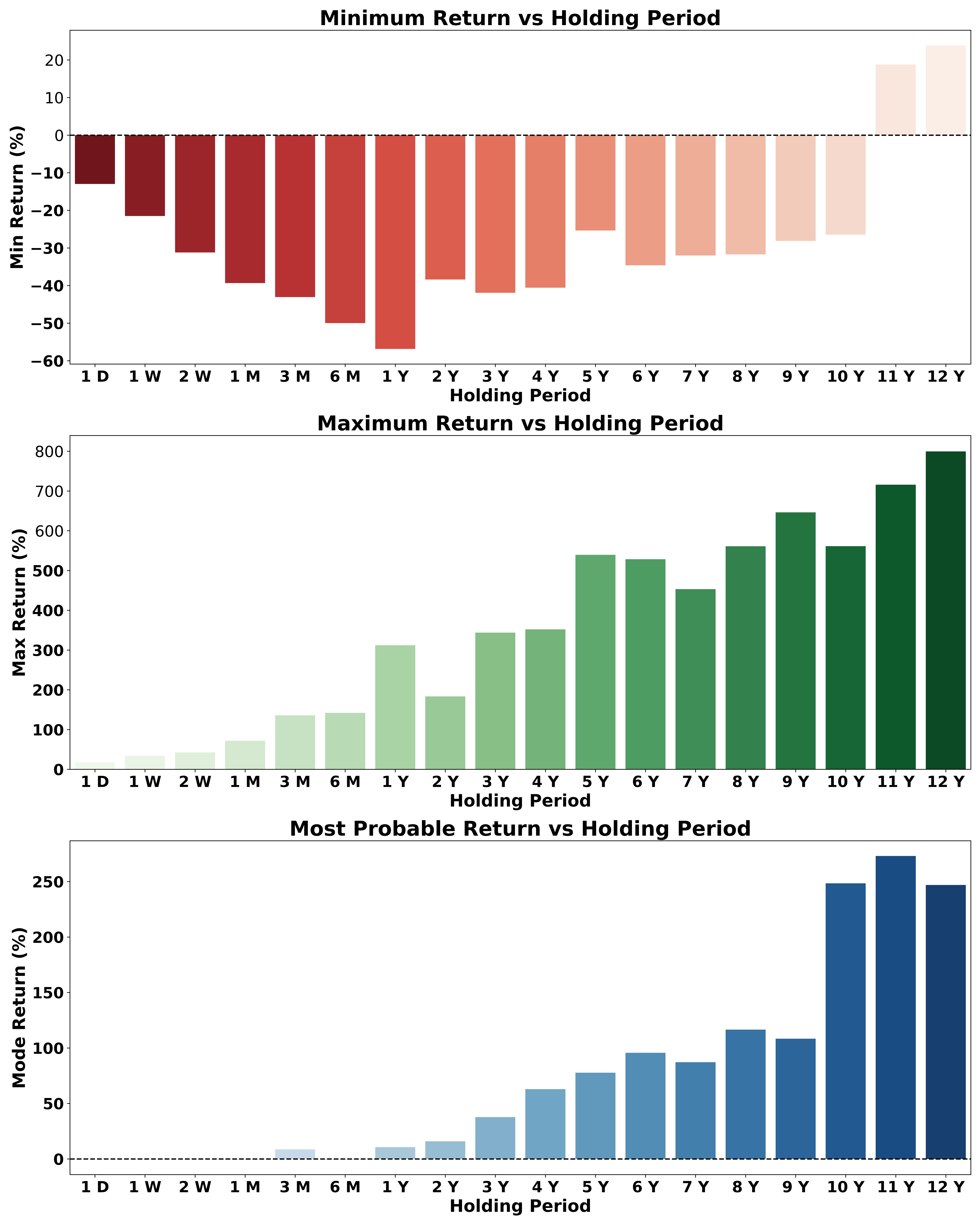}
     \caption{Summary of Minimum, Maximum, and Mode Returns for Nifty 50 Across Investment Horizons (1990--2024).}
     \label{fig:Nifty50mmmreturn90}
 \end{figure}

To synthesize these insights, Figure~\ref{fig:Nifty50mmmreturn90} presents three key statistics across investment horizons: the minimum return (representing the worst-case historical outcome), the maximum return (best-case scenario), and the mode (most probable return). These metrics collectively reveal a compelling time-dependent structure in return distributions. At short horizons—ranging from 1 day to 1 month—returns exhibit limited upside potential and heightened downside risk. For example, the one-day minimum return is \(-12.98\%\), with a mode of just \(0.01\%\), underscoring the equity market’s vulnerability to abrupt shocks and volatility clustering. However, as the investment horizon extends, return distributions become increasingly dispersed and positively skewed. The maximum return expands steadily from \(17.74\%\) at the 1-day horizon to nearly \(800\%\) over 12 years, while the mode turns decisively positive beyond the 3-month horizon and exceeds \(273\%\) at the 11-year mark. These patterns highlight not only the compounding benefits of long-term investing but also the growing asymmetry in return outcomes that favor patient investors.

Yet, this long-run optimism is tempered by a critical observation: the presence of an approximate \textbf{10-year worst-case trapping period}—the longest historical stretch over which investors might have experienced negative or negligible cumulative returns. Specifically, the minimum return remains meaningfully negative even at the 9-year horizon (\(-28.13\%\)) and only turns positive beyond the 10-year mark. This prolonged drawdown risk underscores the significance of entry timing and the hazards of investing at valuation extremes. Thus, while long-term horizons offer substantial return asymmetry and upside potential, they also demand a tolerance for extended periods of under-performance. For investors and policymakers alike, these findings reinforce the need to balance growth expectations with realistic assessments of drawdown risk and investment horizon discipline.

\begin{table}[H]
\centering
\caption{Minimum, Maximum, and Mode CAGR of Nifty 50 Across Multi-Year Holding Periods (1990–2024)}
\label{tab:Nifty50cagr}
\begin{tabular}{|c|r|r|r|}
\hline
\textbf{Holding Period} & \textbf{Min CAGR (\%)} & \textbf{Max CAGR (\%)} & \textbf{Mode CAGR (\%)} \\
\hline
1 Year  & -56.84   & 311.99   & 10.67    \\
2 Year  & -21.50   & 68.36    & 7.66     \\
3 Year  & -16.57   & 64.35    & 11.24    \\
4 Year  & -12.19   & 45.81    & 12.97    \\
5 Year  & -5.68    & 44.93    & 12.18    \\
6 Year  & -6.84    & 35.84    & 11.83    \\
7 Year  & -5.36    & 27.68    & 9.37     \\
8 Year  & -4.66    & 26.63    & 10.14    \\
9 Year  & -3.60    & 25.02    & 8.50     \\
10 Year & -3.03    & 20.79    & 13.29    \\
11 Year & 1.58     & 21.02    & 12.72    \\
12 Year & 1.80     & 20.09    & 10.92    \\
\hline
\end{tabular}
\end{table}

\subsection{Compound Annual Growth Rate (CAGR) Analysis}
\label{subsec:cagr}

While cumulative returns offer a useful measure of realized gains over a given horizon, they do not account for the temporal dimension of compounding. To enable consistent comparisons across varying investment periods, we next examine the compound annual growth rate (CAGR)—a normalized return metric that captures the geometric average annual return required to arrive at a given final value.  Mathematically, CAGR is defined as:
 Mathematically, CAGR is defined as:
\[
\text{CAGR (\%)} = \left( \left( 1 + \frac{R_{\text{abs}}(\%)}{100} \right)^{1/n} - 1 \right) \times 100,
\]
where \( R_{\text{abs}} \) is the absolute return over \( n \) years.

CAGR effectively smooths out the variability of year-to-year returns, offering a clearer and more consistent measure of long-term growth potential. Table~\ref{tab:Nifty50cagr} reports the minimum, maximum, and most probable (mode) CAGR for Nifty 50 investments over holding periods ranging from 1 to 12 years, based on historical data from 1990 to 2024. The evolution of the mode CAGR closely aligns with the pattern observed in mode absolute returns (Figure~\ref{fig:Nifty50mmmreturn90}), revealing a steady upward trend with increasing investment horizon. In the short to medium term, CAGRs exhibit greater volatility and downside risk—highlighted by a minimum CAGR of \(-21.5\%\) over a 2-year period. However, as the horizon lengthens, return variability diminishes and growth outcomes become more stable. Beyond the 10-year mark, even the worst-case CAGR turns positive, while the mode CAGR consistently exceeds \(10\%\), underscoring the historical resilience of Indian equities and reinforcing the importance of investment duration in mitigating risk and enhancing return predictability.

\subsection{Post-1999 Market Dynamics}
\label{subsec:post1999}

\begin{figure}[H]
    \centering
    \includegraphics[width=1.0\textwidth]{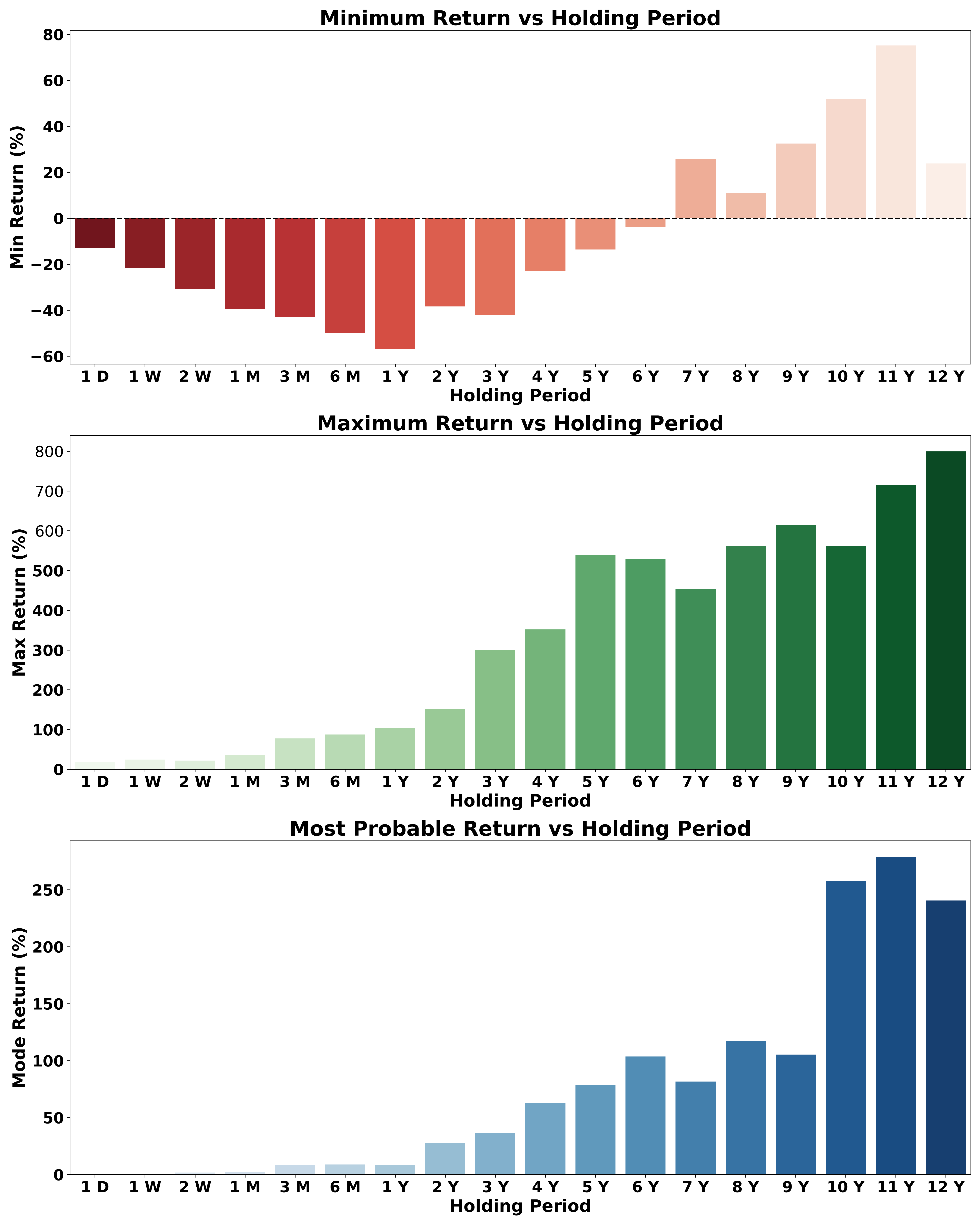}
    \caption{Nifty 50 Multi-Horizon Return Characteristics (1999--2024): Minimum, Maximum, and Mode returns across holding periods.}
    \label{fig:Nifty50return99}
\end{figure}

To further examine how return dynamics have evolved in recent decades, we replicate the multi-horizon CAGR analysis using a narrower dataset covering Nifty 50 performance from January 1, 1999 to December 31, 2024. The full return distributions are presented in Appendix Figure~\ref{fig:Nifty50as99}, with key summary statistics visualized in Figure~\ref{fig:Nifty50return99}. Several notable differences emerge when comparing this period to the broader 1990–2024 sample. 

For instance, the mode of the 1-year return has declined from \(10.67\%\) to \(8.58\%\), reflecting a slight moderation in the most probable short-term outcome for investors. More significantly, the historical worst-case “trapping time”—the longest duration during which cumulative returns remained negative—has compressed markedly. In the earlier sample, a 10-year investment horizon was required to ensure a positive minimum return, whereas in the 1999–2024 window, this threshold drops to just 6 years. This improvement signals stronger market recoveries in the wake of major disruptions, such as the Global Financial Crisis and the COVID-19 pandemic, bolstered by more proactive and coordinated monetary and fiscal responses in the post-reform era.

\begin{table}[H]
\centering
\caption{Minimum, Maximum, and Mode CAGR of Nifty 50 Across Multi-Year Holding Periods (1999--2024)}
\label{tab:Nifty50cagr99}
\begin{tabular}{|c|r|r|r|}
\hline
\textbf{Holding Period} & \textbf{Min CAGR (\%)} & \textbf{Max CAGR (\%)} & \textbf{Mode CAGR (\%)} \\
\hline
1 Year  & -56.84   & 104.43   & 8.58     \\
2 Year  & -21.50   & 58.89    & 13.00    \\
3 Year  & -16.57   & 58.89    & 10.98    \\
4 Year  & -6.35    & 45.81    & 12.97    \\
5 Year  & -2.88    & 44.93    & 12.30    \\
6 Year  & -0.64    & 35.84    & 12.59    \\
7 Year  & 3.32     & 27.68    & 8.90     \\
8 Year  & 1.32     & 26.63    & 10.19    \\
9 Year  & 3.18     & 24.43    & 8.32     \\
10 Year & 4.27     & 20.79    & 13.59    \\
11 Year & 5.23     & 21.02    & 12.88    \\
12 Year & 1.80     & 20.09    & 10.75    \\
\hline
\end{tabular}
\end{table}

CAGR-based metrics further reinforce these observations. As shown in Table~\ref{tab:Nifty50cagr99}, long-term mode CAGRs continue to exceed \(10\%\) in the post-1999 sample, but stabilization occurs earlier, with mode CAGRs clustering around \(12\%\) from the 4-year to 6-year horizons. Additionally, the minimum CAGR turns positive by the 7-year mark—three years earlier than in the 1990–2024 period—highlighting reduced downside asymmetry and accelerated recovery trajectories in more recent market cycles. However, this resilience comes with a trade-off: the maximum CAGR values observed in the 1999–2024 dataset are generally lower than those from the full historical range, suggesting that while downside risk has moderated, the magnitude of extreme upside has also diminished. Overall, the recent era of Indian equity markets exhibits reduced volatility in both directions, signaling enhanced structural stability but also pointing to the maturing nature of growth dynamics.

While distributional analysis reveals the likelihood and asymmetry of returns across time horizons, it does not fully explain the underlying mechanisms driving these outcomes. Return distributions tell us \textit{what} is likely, but not \textit{why} certain patterns emerge or how market behavior evolves. To uncover the structural forces behind price movements—particularly the roles of memory, randomness, and sensitivity to initial conditions—we shift from descriptive statistics to a dynamical systems perspective. This approach enables deeper insights into the informational content, persistence, and nonlinear dynamics inherent in the Nifty 50 index.

\subsection{Analyzing the Complexity of Nifty 50 Returns}
\label{sec:pmf_complexity}
In this context, we complement our empirical return analysis with a study of three widely recognized complexity metrics: Shannon Normalized Entropy (SNE), Generalized Hurst Exponent ($H=H(q=2)$), and the Largest Lyapunov Exponent (LLE). These measures are applied across return horizons from one day to fifteen years time horizons to capture the evolving nature of predictability and structural complexity in the Indian equity market.

\textit{Shannon Normalized Entropy} (SNE) provides a quantitative measure of randomness or disorder in the return time series~\cite{shannon1948, kukreti2020, patra2022}. A value closer to 1 reflects a state of high uncertainty and minimal structure—conditions characteristic of statistical equilibrium—whereas lower values indicate more ordered and potentially predictable behavior. For Nifty 50 returns, SNE exhibits a clear upward trajectory across time horizons: starting at 0.51 for daily returns and increasing steadily to approximately 0.9 for return horizons of 14--15 years. This pattern suggests that short-term market dynamics deviate significantly from equilibrium, influenced by factors such as market microstructure effects, frictions, and behavioral biases. Conversely, the near-saturation of SNE at longer horizons implies a convergence toward equilibrium. This progressive rise in entropy highlights that as the investment horizon extends, the market approaches a state of maximum disorder—consistent with the efficient market hypothesis.

The \textit{Hurst exponent} ($H$), and by extension the \textit{generalized Hurst exponent}, serves as a critical measure of long-range dependence and scaling behavior in complex systems~\cite{shah2024movinghurt, alvarez2008hurst, GomezAguila2022, ZournatzidouFloros2023,barunik2010}. In this framework, values near $H = 0.5$ correspond to systems at statistical equilibrium, exhibiting no memory. Deviations from this value signal broken symmetry and emergent collective dynamics: $H < 0.5$ implies anti-persistence (analogous to mean-reverting or dissipative dynamics), while $H > 0.5$ suggests persistent, trend-reinforcing behavior characteristic of systems near criticality.

For the Nifty 50, estimated $H$ values predominantly lie within the narrow band of 0.5 to 0.56 across all time horizons, suggesting weak but robust persistence consistent with marginally superdiffusive scaling. At the shortest scale (1-day returns), $H \approx 0.0034$, effectively indistinguishable from uncorrelated noise, reflecting high-frequency microstructure effects and short-term randomness. However, as the time horizon increases—a process akin to coarse-graining in renormalization group theory—$H$ converges toward 0.5–0.56. This stabilization implies the emergence of weak memory effects, potentially rooted in macroeconomic cycles, investor herding, or structural characteristics of the market.

Notably, a subtle decline in $H$ beyond the 10-year horizon may indicate transitions between market regimes or underlying structural shifts in the economy, analogous to systems moving away from a critical point. This long-term behavior reflects the dynamic scaling nature of financial markets and aligns with the statistical mechanics perspective that scaling exponents (such as $H$) encapsulate deep structural properties of complex adaptive systems. 

To complement this view, we turn to another key nonlinear study—the \textit{Largest Lyapunov Exponent} (LLE)—which captures a different facet of market complexity: sensitivity to initial conditions and the degree of deterministic chaos. While \( H \) reveals long-memory and persistence patterns, the LLE quantifies the measure of how rapidly initially close trajectories in a system’s phase space move apart exponentially as time progresses~\cite{Tsakonas2022, YanMohammadzadeh2024, eckmann1985, wolf1985, benettin1980, kantz1994}. 

A positive LLE implies the presence of chaos and suggests that the system has ergodic tendencies, as it explores the accessible phase space through divergent trajectories. For Nifty 50 returns, the LLE is approximately 0.5 for daily returns, indicating a high degree of short-term chaoticity and unpredictability—features typically driven by microstructure noise, speculative trading, and behavioral feedback. As the return horizon lengthens, the LLE gradually declines, reaching about 0.23 around the 9–10 year mark. This decreasing trend points to a reduction in chaotic dynamics and a potential weakening of ergodicity at longer horizons, likely due to temporal averaging and the increasing dominance of macroeconomic constraints. Together, the evolving patterns in both \( H \) and LLE highlight a fundamental shift: while short-term financial dynamics are characterized by strong chaotic and ergodic behavior, the long-term structure becomes increasingly regular, governed by slower-moving economic forces and structural limitations.

\begin{figure}[H]
    \centering
    \includegraphics[width=1.0\textwidth]{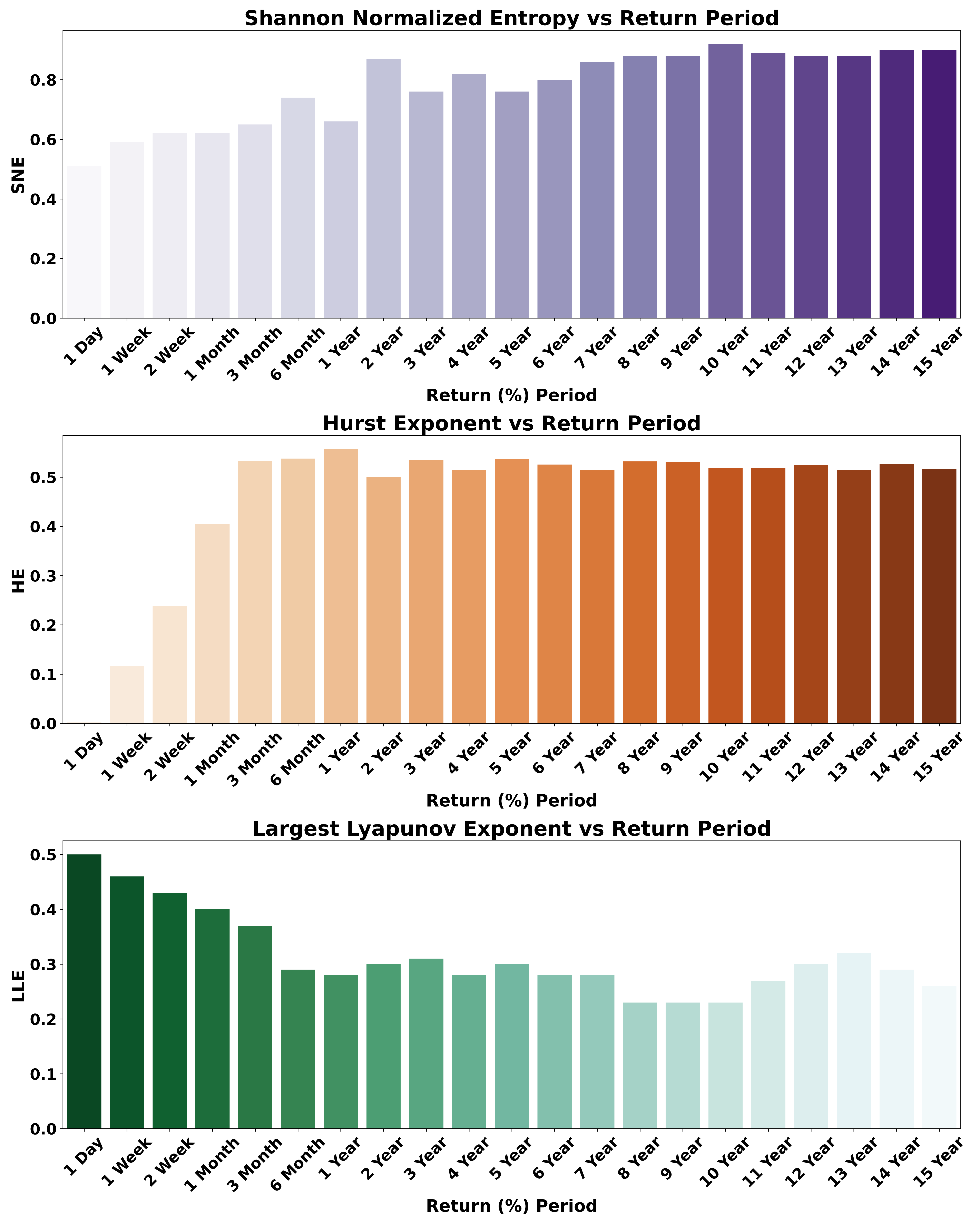}
    \caption{Nifty 50 Return complexity (1999-2024)}
    \label{fig:Nifty50complexity}
\end{figure}

To build a more comprehensive view of market complexity across time horizons, we jointly examine LLE alongside two complementary metrics: Shannon entropy (SNE) and the Hurst exponent ($H$). Figure~\ref{fig:Nifty50complexity} illustrates how all three measures evolve with investment horizon. At short time scales, returns exhibit moderate entropy (SNE~$\approx$~0.5), near-zero memory ($H \approx 0$), and high chaos (LLE~$\approx$~0.5), highlighting a regime dominated by noise and instability. In contrast, long-term returns display high entropy (SNE~$\approx$~0.9), modest persistence ($H \approx 0.5$), and lower chaos (LLE~$\approx$~0.23), suggesting that macroeconomic fundamentals and structural patterns play a greater role over time. Together, these findings reveal a clear transition in market behavior from disorderly and reactive in the short run to more stable and predictable over longer horizons—reinforcing the need for time-scale-aware modeling approaches in financial analysis and strategy design.

\section{Understanding the Complexity of NIFTY 50 P/E ratio}
\label{sec:complexityPE}

Our analysis began with the recognition that elevated P/E ratios alone are insufficient to reliably predict drawdowns in Nifty 50 price movements. This motivated a deeper investigation into the probabilistic structure of returns across multiple time horizons, leading to the study of return distributions through empirical PMFs. We then extended the analysis by exploring the inherent complexity of Nifty 50 returns—quantified using Shannon Normalized Entropy (SNE), the Hurst exponent ($H$), and the Largest Lyapunov Exponent (LLE)—to better understand how randomness, memory, and sensitivity to initial conditions evolve with investment horizon.

These findings emphasize that fully understanding Nifty 50 price behavior—especially outside liquidity-driven episodes—requires joint analysis of return patterns and the statistical and dynamical features of the P/E ratio. While complexity metrics applied to returns shed light on randomness, persistence, and chaos, they offer an incomplete picture without considering how valuation metrics like the P/E ratio evolve over time. Investigating the probability mass function and complexity features of the P/E ratio can reveal underlying structural patterns, behavioral thresholds, and nonlinear feedback mechanisms often missed by static approaches. This dual-layered framework—built upon both return dynamics and valuation-based complexity—offers a more integrated and forward-looking perspective for interpreting market behavior.

\subsection{Probabilistic Structure of the Nifty 50 P/E Ratio }
\label{sec:PEdist}

\begin{figure}[t]
\centering
\includegraphics[width=1.0\linewidth, keepaspectratio]{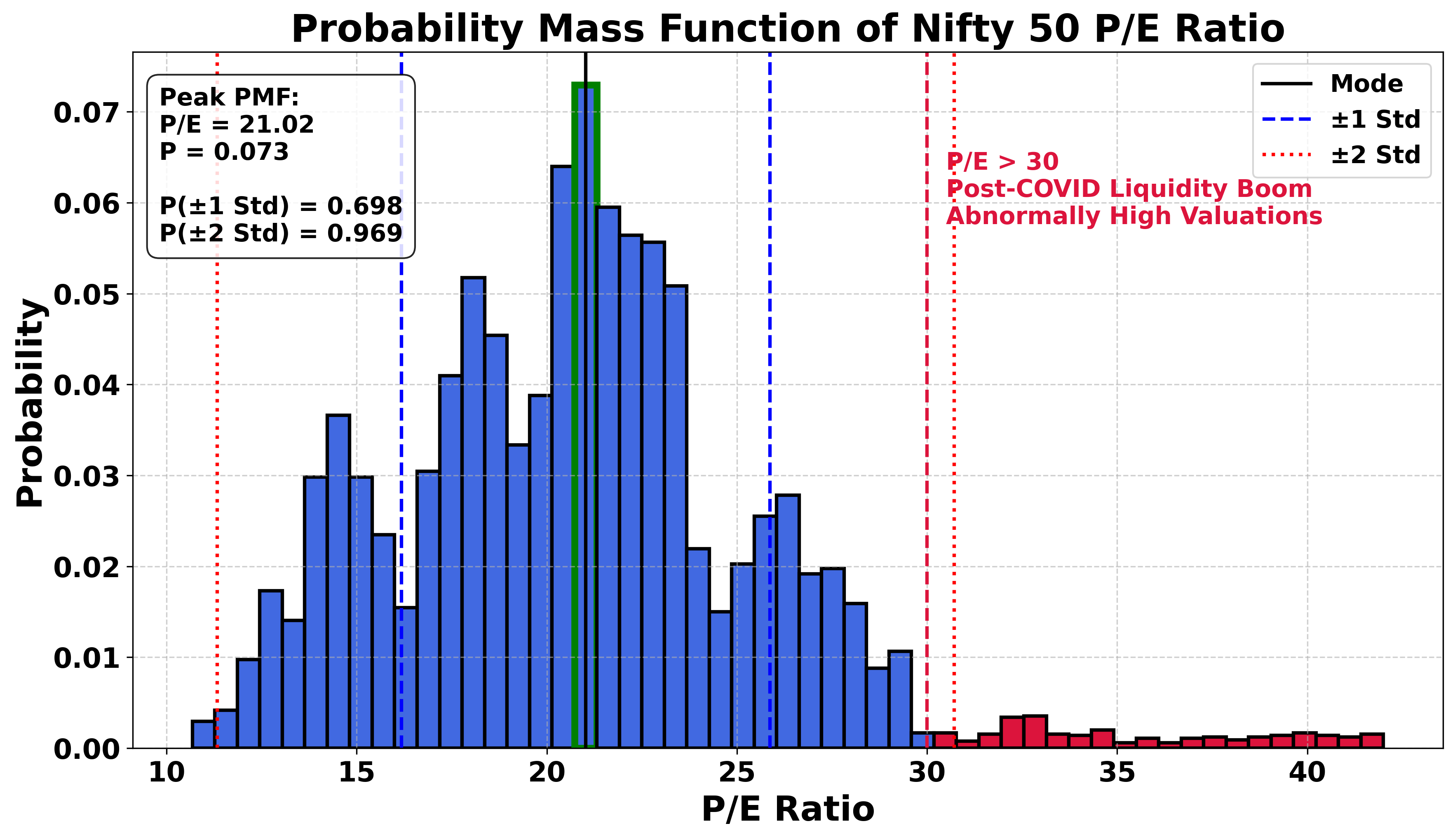}
\caption{PMF of the Nifty 50 P/E ratio from January 1, 1999, to December 31, 2024.}
\label{fig:Nifty50PE}
\end{figure}

To operationalize this valuation-centric perspective, we begin by analyzing the \textbf{probability mass function (PMF)} of the Nifty 50 Price-to-Earnings (P/E) ratio. The dataset consists of daily observations from \textbf{January 1, 1999, to December 31, 2024}, offering a comprehensive view across multiple market cycles.

For robust and adaptive estimation of the PMF structure, we employ the Freedman–Diaconis rule~\cite{scott1979, freedman1981, rudemo1982, birge2006} to determine bin widths. This method ensures that the binning is sensitive to both sample size and variability, thus preserving important distributional features without over-smoothing.

Figure~\ref{fig:Nifty50PE} reveals that the empirical PMF has a well-defined mode at \textbf{21.02}, with a standard deviation of \textbf{4.85}. These distributional characteristics offer more than descriptive insight—they provide a statistical basis for defining valuation zones. By linking P/E ratio levels to their observed probabilities, we establish a data-driven and context-sensitive framework for interpreting market valuation. These statistics form the basis for probabilistically grounded valuation zones. Specifically:

\begin{itemize}
    \item The \textbf{±1$\sigma$ interval} (16.18 to 25.87) captures approximately \textbf{69.82\%} of historical P/E observations.
    \item The broader \textbf{±2$\sigma$ interval} (11.33 to 30.71) accounts for roughly \textbf{96.88\%} of values.
    \item The cumulative probability for observing P/E below 26 is \textbf{86.62\%}, and below 30.71 is \textbf{97.17\%}.
\end{itemize}

To aid interpretability, vertical markers are overlaid on the PMF: the mode is indicated by a solid black line; blue and red dotted lines represent the ±1$\sigma$ and ±2$\sigma$ bands, respectively; and the highest PMF bar is highlighted in green to emphasize the most probable valuation zone.

Valuations exceeding \textbf{30}—such as those during the post-COVID liquidity surge   (2020 – 2021) —  lie well beyond the 2$\sigma$ threshold, occurring in only about \textbf{3\%} of the sample. These rare episodes of extreme overvaluation coincided with negative GDP growth and are indicative of significant dislocation from fundamental value, driven largely by exogenous liquidity influxes. Investing during such periods entails heightened downside risk, particularly under conditions of mean reversion or tightening monetary policy.

From a practical standpoint, these distributional insights inform several actionable strategies:

\begin{enumerate}
    \item \textbf{Valuation-Based Entry Timing:} Entering equity positions when the P/E is below \textbf{26} aligns with historical norms, reducing exposure to elevated valuation risk. Approximately \textbf{86\%} of past trading days fall within this range, suggesting statistically favorable entry points.

    \item \textbf{Dynamic Rebalancing:} When the P/E crosses above the +1$\sigma$ or +2$\sigma$ thresholds, a shift toward defensive allocations—such as liquid funds, gold ETFs, or short-duration debt—may be prudent. Conversely, during low P/E regimes within or below the 1$\sigma$ band, overweighting equities becomes risk-reward optimal.

    \item \textbf{Probabilistic Asset Allocation:}
    \begin{itemize}
        \item P/E $<$ 16: Aggressive equity positioning
        \item 16 $\leq$ P/E $<$ 26: Balanced allocation
        \item 26 $\leq$ P/E $<$ 30: Cautionary exposure
        \item P/E $\geq$ 30: Defensive stance; consider hedging
    \end{itemize}

    \item \textbf{Valuation-Aware SIP Optimization:} Systematic Investment Plans (SIPs) can be enhanced by dynamically adjusting equity allocations based on current P/E relative to the PMF, increasing contributions during undervaluation phases and tapering during speculative extremes.
\end{enumerate}

Overall, the empirical PMF of the Nifty 50 P/E ratio offers a robust, data-driven foundation for valuation-aware decision-making. Unlike fixed historical averages or static thresholds, this probabilistic framework contextualizes current market valuations within their long-term distribution. It enables informed navigation of valuation regimes, liquidity distortions, and speculative excesses—ultimately supporting disciplined and adaptive portfolio strategies. Incorporating such probabilistic tools into \textbf{dynamic asset allocation models}, \textbf{market-timing overlays}, and \textbf{risk-sensitive frameworks} can significantly improve investor outcomes and help mitigate behavioral biases.

\subsection{Monthly Probability Structure of Nifty 50 P/E Ratio }
\label{sec:PEdistm}

To deepen our understanding of valuation behavior across different temporal scales, we extend our analysis from daily to monthly data. Using daily Nifty 50 P/E ratio values from January 1, 1999, to December 31, 2024, we assign each observation to its corresponding calendar month and compute discrete, month-wise empirical PMFs. The resulting monthly PMFs reveal several key insights into valuation dynamics:

First, the majority of months display a peak concentration around P/E values of 20--23, indicating a historically stable valuation zone. However, notable deviations occur during the 2020--2021 period, where P/E ratios frequently exceed 30 across multiple months---including May, July, October, and December---despite concurrent GDP contraction. These anomalies reflect a liquidity-driven post-COVID rally, decoupled from macroeconomic fundamentals.

Further analysis uncovers seasonal features: months such as March, June, and October exhibit broader P/E distributions with multi-modal structures, hinting at heightened valuation uncertainty. These patterns may coincide with recurring events such as fiscal closings, earnings releases, or global financial developments. Conversely, months like February and April show pronounced tail behavior, indicating occasional but significant low or high valuation extremes.

This month-wise probabilistic mapping offers several actionable insights for investment strategies:

\begin{itemize}
    \item \textbf{Tactical Rebalancing:} Monitoring the position of the current P/E ratio within its historical monthly PMF enables timely risk management. For instance, when P/E exceeds the 90th percentile for a specific month, shifting toward defensive assets becomes statistically justified.
    
    \item \textbf{Risk Mitigation:} Favoring equity allocations during months when current P/E levels fall within or below the historical modal zone can improve expected return outcomes by aligning with favorable valuation regimes.
    
    \item \textbf{Multi-Asset Positioning:} Persistently elevated valuations---especially beyond 2$\sigma$ thresholds---may warrant temporary allocation to safer instruments such as liquid funds, gold ETFs, or short-duration bonds.
\end{itemize}

Importantly, the observed monthly patterns emphasize the complexity of market valuation behavior. The temporal evolution of P/E ratios is not linear; sudden jumps, regime shifts, and long stretches of elevated or suppressed valuations are evident. Months like March and October often show chaotic switching between valuation regimes, consistent with the behavior of complex adaptive systems influenced by feedback loops, investor sentiment, and macroeconomic shocks.

Although a general mean-reverting tendency exists in the long-term distribution of the P/E ratio, we also observe extended periods of deviation from historical norms---such as the prolonged overvaluation from 2020 to late 2021. This combination of short-term chaos and intermediate-term trends supports a regime-switching interpretation of market behavior.

Overall, the monthly PMF analysis of the Nifty 50 P/E ratio provides a probabilistic, seasonally aware framework for valuation assessment. By recognizing recurring structural features and their deviations, investors can make more informed, data-driven decisions regarding entry timing, risk exposure, and allocation adjustments.

 \subsection{Nonlinearity and Complexity of Nifty 50 P/E Ratio}
\label{sec:pe_complexity}

Having established the probabilistic distribution and temporal  patterns of the Nifty 50 P/E ratio across daily and monthly scales, we now turn to a more fundamental question: \textit{To what extent does the P/E ratio exhibit underlying nonlinearity, memory, and chaotic structure}? Given the nonlinearities already observed in the PMFs—including multimodal behavior, seasonal asymmetries, and regime shifts—this section systematically evaluates the complexity of the P/E time series using tools from information theory, fractal analysis, and dynamical systems theory.

\subsubsection{Entropy Measures: Probing the Nature of Complexity}
\label{sec:entropy_analysis}

To initiate a rigorous exploration of the complexity inherent in the Nifty 50 P/E ratio time series, we begin with entropy-based measures, which quantify the amount of disorder, unpredictability, and information content in a system. However, complexity in financial time series is multifaceted—arising from nonlinear dependencies, multiscale structures, stochastic fluctuations, and potential deterministic chaos. No single entropy measure can fully capture all these characteristics.

To address this, we employ a diverse set of entropy metrics, each designed to highlight different dimensions of complexity:

\begin{itemize}
    \item \textbf{Shannon Entropy} provides a baseline estimate of average information content, assuming probabilistic independence and stationarity~\cite{shannon1948, kukreti2020, patra2022}.
    
    \item \textbf{Tsallis Entropy} generalizes Shannon entropy by incorporating non-extensivity, making it sensitive to long-range interactions, heavy tails, and deviations from equilibrium—common features in financial data~\cite{antoniades2021tsallis,  Devi2021,  Tian2023}.
    
    \item \textbf{Sample Entropy} assesses the regularity and predictability of fluctuations, particularly useful for identifying short-term deterministic structure in noisy signals~\cite{Pincus1991,  RichmanMoorman2000,  Zunino2010,  Ahmad2013,  Nikbakht2018,  Omidvarnia2019,   Alkan2023}.
    
    \item \textbf{Permutation Entropy} captures the temporal ordering of values, enabling detection of dynamical complexity and distinguishing between stochastic and chaotic behaviors~\cite{Zunino2018, Alves2020,  ChenMaFuLi2023,   Obanya2024}.
\end{itemize}

By combining these complementary entropy measures, we aim to construct a more holistic and robust picture of the informational and dynamical complexity embedded in the P/E ratio time series. The computed results are presented in Table~\ref{tab:entropy_results}, providing multi-angle evidence on the stochastic and potentially deterministic nature of the system.

\begin{table}[h!]
    \centering
    \caption{Entropy Measures of Nifty 50 P/E Ratio}
    \label{tab:entropy_results}
    \begin{tabular}{lll}
        \toprule
        \textbf{Library} & \textbf{Entropy Type}  & \textbf{Normalized Value} \\
        \midrule
        SciPy   & Shannon Entropy            & 0.86 \\
        Manual  & Tsallis Entropy ($q=0.1$)  & 0.92 \\
        Manual  & Tsallis Entropy ($q=2$)     & 0.98 \\
        nolds   & Sample Entropy             & 0.10 \\
        antropy & Permutation Entropy        & 0.94 \\
        \bottomrule
    \end{tabular}
\end{table}

The computed entropy values offer a multi-dimensional view of the complexity embedded in the Nifty 50 P/E ratio time series. Each entropy metric captures different facets of randomness, structure, and dynamical behavior.

The normalized Shannon entropy is 0.86, indicating high—but not maximal—uncertainty in the P/E series. This suggests substantial variability in the data, yet also implies some degree of structure or predictability preventing the entropy from reaching 1.0. 

To refine this picture, we consider the Tsallis entropy, which allows sensitivity to the statistical weight of rare versus frequent events through the entropic index $q$. For $q = 0.1$, the entropy value of 0.92 emphasizes the influence of rare events, while for $q = 2$, the higher value of 0.98 highlights the dominance of frequent structures. This divergence across $q$ values reflects the coexistence of persistent valuation regimes and sporadic, large deviations—characteristics typical of complex financial systems with fat-tailed distributions and multiscale dynamics.

Sample entropy, with a notably low value of 0.10, indicates that the P/E series exhibits significant temporal regularity. Unlike Shannon or Tsallis entropy, which operate primarily on distributional characteristics, sample entropy directly assesses the recurrence of patterns over time. The low value here points to underlying deterministic or structured dynamics, rather than pure randomness.

Permutation entropy, computed as 0.94, adds further insight into the temporal structure by measuring the complexity of ordinal patterns. The high value suggests a rich, possibly chaotic ordering in the time series evolution—consistent with nonlinear dynamical behavior observed in complex systems.

Taken together, these results reveal that the Nifty 50 P/E ratio is neither entirely random nor fully deterministic. Instead, it exhibits a nuanced mix of high variability, long-range dependencies, short-term regularities, and complex temporal patterns. This validates the use of diverse entropy measures as essential tools for uncovering the layered complexity of financial time series.

While entropy measures are powerful tools for quantifying complexity, they primarily capture a static view of the information content and short-term unpredictability in a time series. They do not explicitly characterize how fluctuations evolve across different time scales or how memory effects vary depending on the magnitude of fluctuations. In financial systems, where small and large deviations may exhibit fundamentally different dynamics, such scale-dependent behavior is critical to understanding systemic complexity.

To investigate these aspects, we turn to the \textbf{Generalized Hurst Exponent} $H(q)$, which extends the classical Hurst analysis to capture multiscaling behavior and fluctuation heterogeneity. Specifically, by analyzing $H(q)$ for different values of the moment order $q$, we can assess whether the P/E ratio dynamics follow a simple scaling law or exhibit multifractality—indicative of a complex structure where persistence and memory vary with the intensity of fluctuations.

\subsubsection{Multifractal Behavior via Generalized Hurst Exponents}
\label{sec:hurstexp}

To probe the scaling properties and memory characteristics of the P/E ratio, we compute the Generalized Hurst Exponent $H(q)$ for $q = 1$ to $5$. A decreasing $H(q)$ with increasing $q$ signals multifractality—heterogeneous scaling behavior across fluctuation magnitudes. This approach allows us to quantify how the persistence or anti-persistence of the time series depends on the size of variations, thus offering a deeper, more refined view of temporal dependencies beyond what entropy measures alone can reveal.

\begin{table}[htbp]
\centering
\caption{Generalized Hurst Exponents for Nifty 50 P/E Ratio}
\label{tab:GH}
\begin{tabular}{|c|c|}
\hline
Order $q$ & Generalized Hurst Exponent $H(q)$ \\
\hline
1 & 0.5573 \\
2 & 0.5351 \\
3 & 0.4899 \\
4 & 0.4300 \\
5 & 0.3768 \\
\hline
\end{tabular}
\end{table}

The monotonic decline of $H(q)$, shown in Table~\ref{tab:GH}, confirms multifractality. Specifically:
\begin{itemize}
    \item $H(1) = 0.56$ suggests persistent behavior in small fluctuations,
    \item $H(2) = 0.53$ indicates modest long-range dependence,
    \item $H(3$–$5) < 0.5$ implies anti-persistent or mean-reverting behavior in larger fluctuations.
\end{itemize}

This multifractal nature suggests that volatility and memory in the P/E ratio are scale-dependent—likely driven by heterogeneous investor reactions, policy shocks, and liquidity cycles~\cite{shah2024movinghurt, alvarez2008hurst, GomezAguila2022, ZournatzidouFloros2023, mandelbrot1997variation, mantegna1999introduction, gopikrishnan1999scaling, plerou1999scaling, micciche2003scaling, lux2004detecting, christensen2012multiscale, sen2016time, luskin2017multi}.

While the multifractal analysis provides valuable insights into the scale-dependent behavior and memory effects in the P/E ratio time series, it does not fully capture the deterministic structure or complexity of the underlying dynamics. In particular, multifractality alone cannot distinguish whether the system is driven purely by stochastic processes or exhibits deterministic chaos.

To address this, we investigate the system's sensitivity to initial conditions and explore its underlying nonlinear dynamics through the \textbf{Lyapunov spectrum}. Lyapunov exponents measure the average exponential rate at which nearby trajectories in phase space diverge or converge, providing a direct and quantitative indicator of chaotic dynamics.

\subsubsection{Lyapunov Spectrum: Evidence of Chaos}
\label{sec:lyapunovexp}

We further explore the nonlinear dynamical structure of the P/E ratio by computing the full Lyapunov spectrum using a 5-dimensional embedding. The results are summarized in Table~\ref{tab:lyap_spectrum}. The presence of two positive Lyapunov exponents confirms the existence of low-dimensional chaos, indicating the system’s sensitivity to initial conditions and complex underlying dynamics.

Pesin’s Theorem provides a direct link between deterministic chaos and entropy in smooth dynamical systems. Specifically, it states that for a dynamical system with an absolutely continuous invariant measure and a Lyapunov spectrum $\{\lambda_i\}$, the Kolmogorov–Sinai (KS) entropy $h_{KS}$ is equal to the sum of the positive Lyapunov exponents~\cite{pesin1977,young1982}:
\[
h_{KS} = \sum_{\lambda_i > 0} \lambda_i.
\]
This identity holds under the assumption that the system is ergodic and differentiable almost everywhere. In chaotic systems, $h_{KS}$ quantifies the average rate of information loss and thus characterizes the level of unpredictability.

In the case of the Nifty 50 P/E ratio, the computed Lyapunov spectrum (Table~\ref{tab:lyap_spectrum}) yields two positive exponents:
\[
\lambda_1 = 0.30, \quad \lambda_2 = 0.11.
\]
Applying Pesin’s formula gives an approximate KS entropy of:
\[
h_{KS} \approx \lambda_1 + \lambda_2 = 0.41.
\]

This positive $h_{KS}$ value confirms the presence of low-dimensional chaos in the P/E ratio time series. It implies that the system generates approximately $0.41$ nats (natural units of information) per time step, reflecting a moderate but persistent level of dynamical complexity and irreversibility. The nonzero entropy rate highlights the limits of precise predictability and supports the use of nonlinear and entropy-aware frameworks in modeling valuation dynamics and return structures.

\begin{table}[h!]
\centering
\caption{Lyapunov Spectrum of Nifty 50 P/E Ratio}
\label{tab:lyap_spectrum}
\begin{tabular}{|c|c|}
\hline
\textbf{Lyapunov Exponent} & \textbf{Value} \\
\hline
$\lambda_1$ & 0.30 \\
$\lambda_2$ & 0.11 \\
$\lambda_3$ & -0.08 \\
$\lambda_4$ & -0.28 \\
$\lambda_5$ & -0.74 \\
\hline
\end{tabular}
\end{table}

Additionally, the Kaplan–Yorke (or Lyapunov) dimension, $D_{KY}$, provides an estimate of the attractor’s fractal dimension based on the Lyapunov spectrum. It is defined as~\cite{Kaplan1979, Ott2002, Wu2021,   Kachhia2023,  Goyal2024}:

\[
D_{KY} = j + \frac{\sum_{i=1}^{j} \lambda_i}{|\lambda_{j+1}|},
\]
where \(j\) is the largest integer such that
\[
\sum_{i=1}^{j} \lambda_i \geq 0 \quad \text{and} \quad \sum_{i=1}^{j+1} \lambda_i < 0.
\]

Using the Lyapunov exponents from Table~\ref{tab:lyap_spectrum}:
\[
\lambda_1 = 0.30, \quad \lambda_2 = 0.11, \quad \lambda_3 = -0.08, \quad \lambda_4 = -0.28, \quad \lambda_5 = -0.74,
\]
we compute the partial sums:
\[
S_1 = 0.30, \quad S_2 = 0.41, \quad S_3 = 0.33, \quad S_4 = 0.05, \quad S_5 = -0.69.
\]

Since \( S_4 \geq 0 \) but \( S_5 < 0 \), we set \( j = 4 \), yielding:
\[
D_{KY} = 4 + \frac{0.05}{0.74} \approx 4.07.
\]

This result indicates a low-dimensional chaotic attractor with fractal structure, reinforcing the presence of deterministic chaos and multifractal scaling in the Nifty 50 P/E ratio dynamics.

While the presence of chaos, as evidenced by positive Lyapunov exponents and fractal attractors, confirms sensitive dependence on initial conditions, it does not directly tell us how the P/E ratio dynamics influence or relate to the actual price movement of the Nifty 50 index. In other words, chaotic behavior alone does not quantify the degree to which past values of the P/E ratio contribute information about future index returns.

To bridge this gap, we turn to \textbf{mutual information}—a tool that can capture both linear and nonlinear dependencies. In particular, we compute the lagged \textbf{Normalized Mutual Information (NMI)} between the P/E ratio and Nifty 50 price movement to assess how much predictive information is retained over time.

\subsubsection{Normalized Mutual Information: \emph{amount of information} that the P/E ratio provides about the index’s price changes}
\label{sec:mutualinfo}

Investors and analysts constantly seek indicators that can help predict stock market movements. Among these, the \textbf{Price-to-Earnings (P/E) ratio} of the Nifty 50 index is widely followed as a gauge of market valuation and investor sentiment. But how much does the P/E ratio actually tell us about the future price movement of the Nifty 50?

To answer this, we need a way to measure the \emph{amount of information} that the P/E ratio provides about the index’s price changes. Simply observing correlations or trends may not capture the full picture, especially if the relationship is complex or nonlinear.

This is where \textbf{mutual information} comes in --- a powerful tool from information theory that quantifies the degree of dependence between two variables, regardless of the nature of their relationship~\cite{Fiedor2014PhysA,   Fiedor2014PRE,  Wang2017, Goh2018, Guo2018}.

By studying the mutual information between the Nifty 50’s P/E ratio and its price movement, we can uncover how much knowing the P/E ratio actually reduces our uncertainty about the index’s future direction. This insight helps investors understand the true predictive power of the P/E ratio and refine their decision-making strategies.

In essence, exploring this relationship moves us beyond intuition and conventional metrics, allowing for a deeper, data-driven understanding of market dynamics.

\begin{figure}[H]
    \centering
    \includegraphics[width=0.75\textwidth]{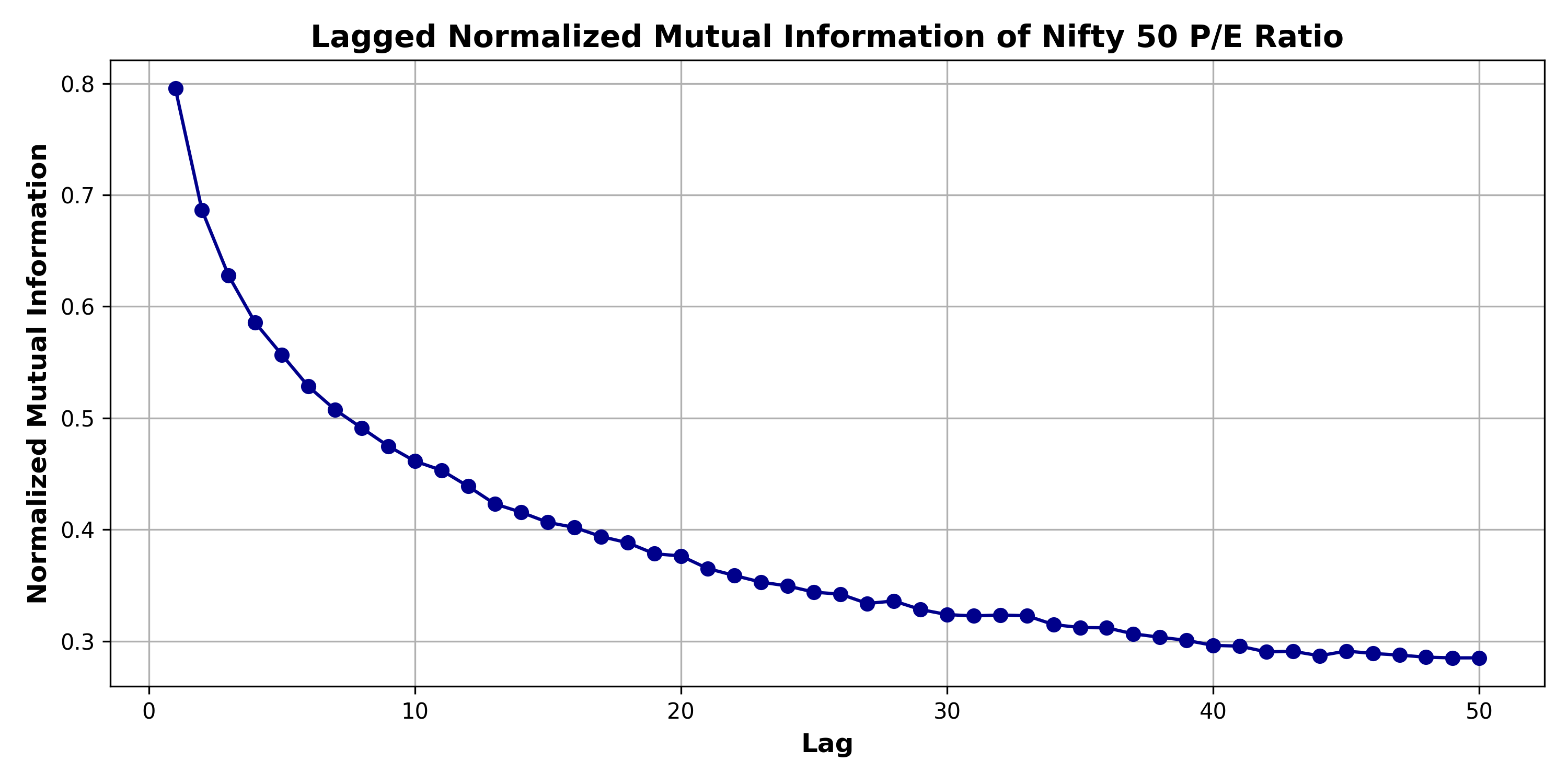}
    \caption{Lagged Normalized Mutual Information of Nifty 50 P/E Ratio}
    \label{fig:nmi_plot}
\end{figure}

To investigate the mutual information between them, we compute the lagged Normalized Mutual Information (NMI) up to lag 50 (see Figure~\ref{fig:nmi_plot}). 

The normalized mutual information (NMI) between the P/E ratio and the Nifty 50 price movement at lag 1 is 0.795, indicating a strong immediate information flow from the P/E ratio to price changes. This information transfer remains significant, with NMI values staying above 0.4 for up to 15 lags, reflecting a sustained influence of past P/E ratios on future price movements. Even at lag 50, the NMI remains around 0.28, which is well above what would be expected by chance, demonstrating a persistent and long-lasting dependency between the two variables.

Such enduring information flow, combined with the low sample entropy and chaotic dynamics observed in the P/E ratio series, suggests that the relationship between the P/E ratio of Nifty 50 and it's price movement is neither memoryless nor purely random. Instead, it exhibits a structured, complex dependency that can be leveraged to better understand and potentially predict market behavior.

\subsubsection{Implications and Modeling Considerations}
\label{sec:Imp_mod_cons}
The convergence of results across multiple domains—information theory, fractal analysis, and chaos theory—strongly establishes the Nifty 50 P/E ratio as a complex, nonlinear system with both deterministic and stochastic elements. The system displays:
\begin{itemize}
    \item High entropy and informational complexity,
    \item Multifractality across scales,
    \item Sensitivity to initial conditions (chaos),
    \item Long-range memory and regime switching behavior.
\end{itemize}

These features challenge traditional linear models and underscore the necessity for nonlinear, multifractal, and chaos-informed frameworks for modeling market valuations. The presence of structured complexity also suggests potential predictability windows, which may be exploited through advanced machine learning, nonlinear filtering, or regime-dependent asset allocation strategies.

In conclusion, the P/E ratio is not merely a static valuation indicator—it is a dynamic, evolving entity governed by complex market forces and behavioral feedback. Properly modeling its evolution requires embracing the full toolkit of nonlinear time series analysis.

\section{Conditional Return Analysis of Nifty 50 Based on P/E Ratio Bands}
\label{sec:returngivenPEratio}

While prior sections establish that the Nifty 50 P/E ratio exhibits nonlinear, chaotic, and multifractal properties, its utility as a predictor of future market behavior remains an open question. This section addresses this issue by examining the informational and causal relationships between the P/E ratio and subsequent index returns using information-theoretic tools, followed by a conditional probabilistic framework that explores how valuation levels influence long-term return outcomes.

\subsection{Mutual Information and Transfer Entropy: Nonlinear Dependence and Causality}
\label{sec:MI_TE}
We first evaluate the mutual dependence between the Nifty 50 P/E ratio and next-day returns using \textit{Mutual Information} (MI). The estimated MI value is approximately \textbf{0.0405}, suggesting a weak but non-zero relationship. While this low value implies that the P/E ratio contains limited predictive information for short-term price changes, it is consistent with the characteristics of financial time series, which are inherently noisy and influenced by a multitude of factors including macroeconomic indicators, market sentiment, and exogenous shocks.

Importantly, the presence of even a modest MI value supports the hypothesis that valuation levels may exert some influence on immediate return behavior, albeit insufficient for standalone forecasting. This highlights the potential for the P/E ratio to function as a weak signal that may become more informative when combined with other fundamental or technical indicators in a multivariate or regime-based framework.

To further investigate the dynamic relationship between valuation and market behavior, we compute the \textit{Transfer Entropy} (TE) from the P/E ratio to next-day returns. While Mutual Information (MI) effectively quantifies the strength of dependence between two variables, it is symmetric in nature and does not account for the direction of information flow. In contrast, TE is an asymmetric measure rooted in information theory that captures both linear and nonlinear dependencies while explicitly distinguishing the source and target of information transfer~\cite{DimpflPeter2013, DimpflPeter2014, Harre2014, KorbelJiangZheng2019, YaoLi2020}.

This directional property is particularly valuable in financial time series, where understanding whether past valuation metrics (such as the P/E ratio) exert a predictive influence on future price movements is critical for both theoretical modeling and practical forecasting. By applying TE, we aim to determine whether the P/E ratio contains actionable forward-looking information that could inform short-term return dynamics, beyond what symmetric dependence measures can reveal.  For a history length of $k=1$, the TE from the P/E ratio to returns is estimated to be \textbf{0.0306}, while the reverse TE—from returns to P/E—is markedly lower at \textbf{0.0086}. This asymmetry indicates a stronger influence of valuation on subsequent returns than vice versa, suggesting that the P/E ratio serves as a causal precursor, albeit weakly, to short-term price changes. The lower TE in the reverse direction implies limited feedback from price movements to valuation within such short horizons.

Together, MI and TE findings reinforce the notion that the P/E ratio, though not a strong short-term predictor, contains subtle nonlinear dependencies that could be exploitable under specific market conditions or in combination with other variables.

\subsection{Conditional Return Distributions of Nifty 50  Based on it's P/E ratio Bands}
\label{sec:cond_return_givenPE}

Motivated by the information-theoretic evidence—specifically, the mutual information (MI) and transfer entropy (TE) findings that reveal subtle, nonlinear dependencies between the P/E ratio and future returns—we shift our focus from short-term predictability to the broader question of how initial valuation levels shape the distribution of long-term outcomes. To this end, we construct \textit{conditional probability mass functions} of Nifty 50 returns across multiple holding periods (1–7 years), conditioned on discrete P/E ratio bands~\cite{Ghysels2020, NoretsPelenis2022, MaynardShimotsuKuriyama2023, marsili1999scaling, Yan2022_PEN}. This framework enables us to assess whether valuation-dependent structures emerge more clearly at longer horizons—offering practical insights for long-term investors. We limit our analysis to holding periods of up to 7 years because beyond this horizon, the probability of negative returns effectively falls to zero, rendering further horizons unnecessary for estimating meaningful reward-risk ratios.

Hence, the P/E spectrum is discretized into 1-point intervals ranging from 10 to 31 (e.g., 10--11, 11--12, ..., 30--31), with each bin represented by its midpoint (e.g., 10.5, 11.5, ..., 30.5). For each P/E bin, we compute the PMFs of multi-year  returns, resulting in a valuation- and time-conditioned return distribution.  From these PMFs, we derive three summary statistics for each band-duration pair:

\begin{itemize}
    \item \textbf{Probability of Positive Return (PRP)}
    \item \textbf{Probability of Negative Return (NRP)}
    \item \textbf{Reward-Risk Ratio (RRR)}: defined as $\text{PRP}/\text{NRP}$
\end{itemize}

The aggregated results, presented in Appendix Table~\ref{tab:performance-evaluation}, enable a probabilistic evaluation of asymmetries in market outcomes across different valuation levels and investment horizons. In particular, the Reward-Risk Ratio (RRR) serves as a key metric for assessing the balance between upside potential and downside risk, conditioned on initial valuation.

\subsubsection{Reward-Risk Profiles Across P/E Regimes}
\label{sec:RRR}

This analysis uncovers pronounced differences in reward-risk characteristics as the starting P/E ratio varies. By examining how RRR shifts across valuation zones, we gain insight into how market expectations and pricing efficiency evolve across different macro-financial contexts.

\paragraph{P/E $<$ 13: No Risk Zone}
This band is characterized by $\text{NRP} = 0$ across all durations, leading to infinite RRR values. Investors entering at these low valuations faced no observed downside risk historically and enjoyed consistently positive outcomes.

\textit{Implication:} Extremely favorable for entry; ideal for risk-averse investors seeking high confidence in positive long-term returns.

\paragraph{P/E 13--16: Low Risk, Brief Trapping}
Here, NRP rises modestly to 2–7\% for 1-year durations but remains low across longer horizons. RRR values are extremely high (e.g., 174–2866), with breakeven periods typically under 3 years.

\textit{Implication:} Low-risk zone with some short-term volatility, but long-term outcomes are highly favorable.

\paragraph{P/E 16--22: Moderate Risk, 4-Year Trapping}
Negative returns become more common, with NRPs up to $\sim$30\% in the short term. Breakeven periods extend to four years, and RRR values begin to moderate.

\textit{Implication:} Moderate-risk regime; investors may experience multi-year drawdowns before achieving gains.

\paragraph{P/E 22--27: High Risk, 5-Year Trapping}
The short-term NRP increases to as much as 40\%, with breakeven periods averaging five years. While RRRs occasionally appear attractive, the elevated risk level and prolonged drawdown potential undermine short-term investment cases.

\textit{Implication:} Caution advised. Suitable primarily for investors with high risk tolerance and longer horizons.

\paragraph{P/E $>$ 27: Very High Risk, Extended Trapping}
This range shows persistent negative return probabilities even at longer durations. Breakeven periods often exceed six years, and RRR values are highly unstable.

\textit{Implication:} High valuation levels carry significant downside risk and long recovery times, limiting their appeal for most investment strategies.

\subsubsection{Strategic Insights and Practical Applications}

This conditional analysis highlights clear nonlinear and horizon-dependent relationships between valuation and return outcomes. From this, several practical takeaways emerge:

\begin{enumerate}
    \item \textbf{Entry Timing:} Optimal timing occurs when the P/E ratio is below 13, offering nearly risk-free long-term returns.
    \item \textbf{Risk Management:} Higher P/E levels should prompt defensive positioning, hedging, or reduced exposure to avoid extended drawdowns.
    \item \textbf{Horizon Adjustment:} For P/E $>$ 16, short-term investors face meaningful downside risk, while long-term investors may still benefit from valuation mean-reversion.
    \item \textbf{Reward-Risk Optimization:} The RRR metric allows investors to weigh expected payoffs against downside probabilities in each valuation regime.
\end{enumerate}

\subsubsection{Summary and Broader Implications}
\label{sec:summary}
Overall, this analysis underscores a critical insight: the predictive utility of the P/E ratio depends strongly on both the temporal horizon and prevailing valuation regime. While the ratio offers limited short-term forecasting power—as shown by low MI and TE values—it carries considerable weight in shaping multi-year return distributions.

In particular, the results demonstrate that:
\begin{itemize}
    \item \textbf{Low valuations} (P/E $<$ 13) are consistently associated with high probabilities of positive returns and negligible downside risk.
    \item \textbf{Mid-range valuations} (P/E 16--22) carry moderate risk and require a minimum holding period of 3–4 years for favorable outcomes.
    \item \textbf{High valuations} (P/E $>$ 27) entail elevated downside risk and require long horizons ($>$6 years) to offset the initial overvaluation.
\end{itemize}

These findings not only validate long-held principles of value investing but also provide a quantitatively robust, probabilistic framework for tactical asset allocation. The insights can be directly integrated into dynamic portfolio strategies that adapt exposure based on valuation-driven reward-risk asymmetries.

\section{Results and Discussion}
\label{sec:result_discussion}

This section presents a comprehensive empirical and dynamical analysis of the Nifty~50 index (1990--2024), aiming to unify valuation metrics, return distributions, and market complexity under a single interpretive framework. Addressing a key gap in the literature, we condition multi-horizon return dynamics on valuation regimes, while explicitly incorporating nonlinear dependence and complexity features. This approach departs from traditional linear-return modeling by uncovering deeper structural patterns in the behavior of Indian equity markets.

\subsection{Valuation Regimes and Return Asymmetries}

At the heart of this framework is the price-to-earnings (P/E) ratio, a widely used metric in valuation analysis. The Nifty~50 P/E ratio has a modal value of \textbf{21.02} and a standard deviation of \textbf{4.85}; approximately \textbf{69.82\%} of observations lie within its $\pm 1\sigma$ range (16.18–25.87), and \textbf{96.88\%} fall within the $\pm 2\sigma$ range (11.33–30.71). Notably, valuations exceeding 30—accounting for only \textbf{3\%} of observations—were concentrated during the liquidity-driven rally of 2020–2021, largely disconnected from earnings or GDP fundamentals.

Return distributions are highly asymmetric and exhibit strong positive skew. One-day and one-year returns yield reward-to-risk ratios of 1.26 and 5.31, respectively. The modal one-year return is \textbf{10.67\%}, with a \textbf{74\%} historical probability of gain. At longer horizons, modal returns turn consistently positive beyond three months and exceed \textbf{273\%} at eleven years. However, the minimum return remains negative for up to ten years, defining a \textbf{decade-long worst-case trapping period}. Modal CAGR exceeds \textbf{10\%} beyond four years, while worst-case CAGRs become positive only after ten years.

\subsection{Post-Reform Dynamics and Market Resilience}

Segmenting the data post-1999—coinciding with economic liberalization and structural reforms—reveals a distinct improvement in market resilience. During this period, the worst-case breakeven horizon shortens from ten to six years. Modal one-year returns moderate to \textbf{8.58\%}, and minimum CAGRs turn positive by year seven. Modal CAGRs stabilize around \textbf{12\%} for four- to six-year holding periods, suggesting stronger mean-reversion and faster recovery in the reformed market regime.

\subsection{Complexity and Multifractality in Return Dynamics}

To better understand the drivers of these dynamics, we examine market complexity. Entropy-based metrics reveal \textit{structured randomness} in price evolution. Shannon entropy rises from \textbf{0.51} (1-day) to \textbf{0.90} (15-year), indicating an increasing information horizon. The Generalized Hurst exponent, ranging from $H \approx 0.50$ to $0.56$, reflects weak persistence and long-memory characteristics. The Largest Lyapunov Exponent declines from \textbf{0.50} to \textbf{0.23}, indicating reduced chaotic sensitivity at longer horizons.

The P/E ratio itself shows multifractal behavior. Its Hurst exponents—$H(1) = 0.5573$ and $H(2) = 0.5351$—point to persistent small-scale fluctuations and long memory, while $H(3)$ to $H(5) < 0.5$ indicate anti-persistence at broader scales. Low sample entropy (\textbf{0.10}) suggests short-term regularity. The Lyapunov spectrum, with largest exponent $\lambda_1 = 0.30$, secondary exponent $\lambda_2 = 0.11$, KS-entropy of \textbf{0.41}, and attractor dimension near \textbf{4.07}, confirms low-dimensional chaos embedded in valuation dynamics.

\subsection{Information-Theoretic Dependence and Directionality}

We complement the complexity analysis with information-theoretic tools. Normalized Mutual Information (NMI) between the P/E ratio and future returns peaks at \textbf{0.795} (lag 1), remains above \textbf{0.40} through lag 15, and stabilizes at \textbf{0.28} by lag 50, evidencing persistent nonlinear dependence across time.

To explore causality, we compute Transfer Entropy. The information transferred from P/E to returns (\textbf{0.0306}) significantly exceeds the reverse flow (\textbf{0.0086}), suggesting a directional—albeit weak—causal influence from valuation to return outcomes. This directional flow validates the conceptual foundation for conditioning return distributions on valuation regimes.

\subsection{Valuation-Banded Return Profiles}

Conditioning return distributions on valuation bands reveals regime-dependent behavior:

\begin{itemize}
    \item \textbf{P/E < 13}: No historical instances of negative returns at any horizon.
    \item \textbf{P/E 13--16}: Downside risk is minimal (2–7\%) with favorable reward-to-risk ratios.
    \item \textbf{P/E 16--22}: Downside risk increases (up to 30\%), with a four-year breakeven period.
    \item \textbf{P/E 22--27}: Risk intensifies (up to 40\%) and breakeven extends to five years.
    \item \textbf{P/E > 27}: Return instability persists, even at long horizons.
\end{itemize}

These findings highlight the nonlinearity and asymmetry inherent in valuation-return relationships and demonstrate that tail risks and breakeven dynamics are tightly coupled with valuation regimes.

\subsection{Strategic Implications and Contribution}

The results position the P/E ratio as a nonlinear, complexity-rich signal conditioning return distributions over long horizons. While the framework is not designed for short-term return forecasting, it offers valuable insights for long-term portfolio design and downside risk control. It provides a data-driven foundation for dynamic asset allocation policies that align with valuation regimes and complexity diagnostics.

Our contribution lies in integrating conditional return distributions, market complexity, and information-theoretic dependencies within a unified valuation-aware framework. This approach is particularly relevant for practitioners and policymakers operating in emerging markets like India, where traditional linear models often fail to capture structural nonlinearities, regime shifts, and memory effects in asset returns.

\section{Conclusion}
\label{sec:conclusion}

This study develops a unified, distribution-aware, and complexity-informed framework for modeling equity return dynamics in the Indian market, using 34 years of data from the Nifty~50 index (1990--2024). We address a key gap in the literature by demonstrating that a  valuation measure—the price-to-earnings (P/E) ratio—may serve as a nonlinear conditioning variable that probabilistically maps return distributions across investment horizons ranging from days to decades.

Our results reveal pronounced asymmetries in return distributions. For example, one-year returns show a 74\% probability of gain with a modal return of 10.67\%, while long-horizon CAGRs surpass 13\% after a decade. Low valuation regimes (P/E~<~13) historically correspond to zero probability of loss across all horizons, whereas high valuations (P/E~>~27) lead to return instability and extended breakeven periods. These empirical patterns, along with persistent and directional information flow from valuation to returns, underscore the nonlinear and regime-dependent nature of return dynamics in Indian equity markets.

These findings also resonate with prior literature on nonlinear valuation effects and market complexity in developed markets (e.g., \cite{shiller2000irrational, peters1994fractal, cont2001empirical}). By integrating entropy, Hurst exponents, Lyapunov indicators, and information-theoretic measures, our analysis captures long-memory dynamics, weak persistence, and low-dimensional chaos in both price and valuation series. Importantly, transfer entropy confirms valuation’s directional predictive influence on future returns.

Nevertheless, the framework’s short-horizon forecasting ability is limited. External shocks, policy shifts, and structural breaks—especially prevalent in emerging markets—present further challenges. Expanding the scope to include other valuation metrics  and macroeconomic indicators could improve robustness.

Future research may extend this framework to sector-level analysis, incorporate global risk factors, or apply it to other emerging markets to assess generalizability. Such extensions would broaden the applicability of this approach for long-term asset allocation, cycle-aware investing, and risk management.

\section{Acknowledgment}
\label{ack}
 The author gratefully acknowledges the institutional support provided by Birla Institute of Technology and Sciences, Pilani K. K. Birla Goa Campus, which facilitated access to  computational infrastructure, and research guidance. The intellectual environment  at the institute significantly contributed to the development of this work.

\appendix
\label{app}

\section{PMF of Nifty 50}
\label{app0}

\subsection{PMF of Nifty 50 Absolute Return (1990-2024)}
\label{app1}

\begin{figure}[H]
    \centering
    \includegraphics[width=1.0\textwidth]{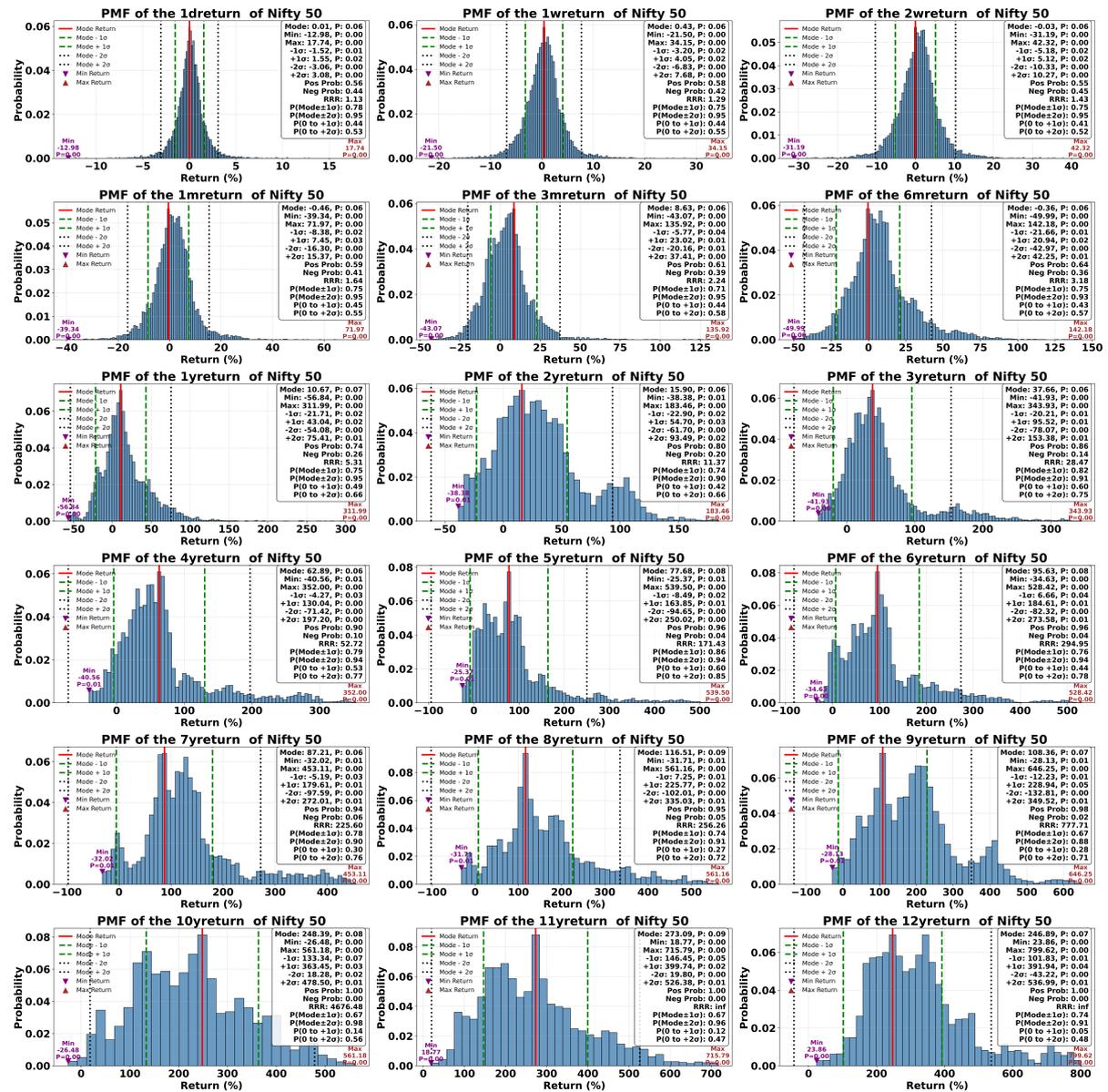}
    \caption{PMF of Nifty 50 Absolute Return (1990-2024)}
    \label{fig:Nifty50as90}
\end{figure}

\subsection{PMF of Nifty 50 Absolute Return (1999-2024)}
\label{app2}
\begin{figure}[H]
    \centering
    \includegraphics[width=1.0\textwidth]{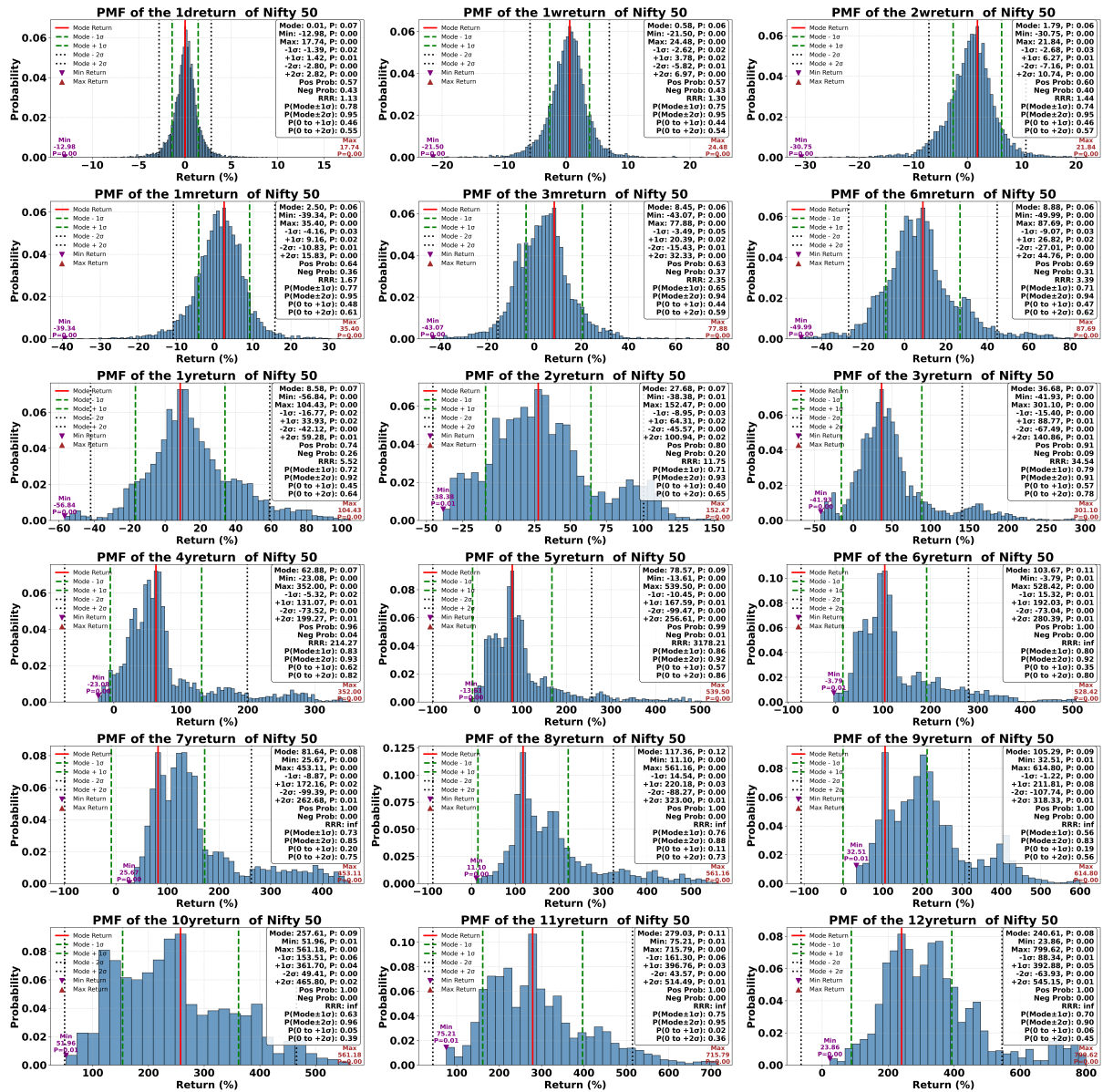}
    \caption{PMF of Nifty 50 Absolute Return (1999-2024)}
    \label{fig:Nifty50as99}
\end{figure}

\subsection{Performance of Conditional Return on  Investment in Nifty 50 given P/E ratio band}
\label{app3}

\begin{longtable}{lllll}
\caption{Performance of Conditional Return on  Investment in Nifty 50 at    P/E ratio band} \label{tab:performance-evaluation} \\
\toprule
\textbf{PE Range} & \textbf{Duration} & \textbf{PRP} & \textbf{NRP} & \textbf{RRR} \\
\midrule
\endfirsthead

\multicolumn{5}{l}{\textbf{Table \ref{tab:performance-evaluation}} (continued)} \\
\toprule
\textbf{PE} & \textbf{Duration} & \textbf{PRP} & \textbf{NRP} & \textbf{RRR} \\
\midrule
\endhead

\bottomrule
\multicolumn{5}{r}{\textit{Continued on next page}} \\
\endfoot

\bottomrule
\multicolumn{5}{p{0.9\textwidth}}{\textit{Note}: PRP = Positive Return Probability, NRP = Negative Return Probability, RRR = Reward-Risk Ratio. $\infty$ denotes cases where NRP = 0.} \\
\endlastfoot

10--11 & 1 Year & 1 & 0 & $\infty$ \\
11--12 & 1 Year & 1 & 0 & $\infty$ \\
12--13 & 1 Year & 1 & 0 & $\infty$ \\
13--14 & 1 Year & 0.98 & 0.02 & 2866.7 \\
13--14 & 2 Year & 1 & 0 & $\infty$ \\
14--15 & 1 Year & 0.93 & 0.07 & 174.4 \\
14--15 & 2 Year & 0.99 & 0.01 & 564 \\
14--15 & 3 Year & 1 & 0 & $\infty$ \\
15--16 & 1 Year & 0.66 & 0.33 & 13.83 \\
15--16 & 2 Year & 0.89 & 0.11 & 44.95 \\
15--16 & 3 Year & 1 & 0 & $\infty$ \\
16--17 & 1 Year & 0.83 & 0.17 & 31.51 \\
16--17 & 2 Year & 0.94 & 0.06 & 62.84 \\
16--17 & 3 Year & 0.98 & 0.02 & 1356.93 \\
16--17 & 4 Year & 0.97 & 0.03 & 448.79 \\
16--17 & 5 Year & 1 & 0 & $\infty$ \\
17--18 & 1 Year & 0.88 & 0.12 & 21.19 \\
17--18 & 2 Year & 0.88 & 0.12 & 24.84 \\
17--18 & 3 Year & 0.89 & 0.11 & 73.63 \\
17--18 & 4 Year & 0.95 & 0.05 & 184.12 \\
17--18 & 5 Year & 1 & 0 & $\infty$ \\
18--19 & 1 Year & 0.83 & 0.17 & 7.83 \\
18--19 & 2 Year & 0.89 & 0.11 & 18.24 \\
18--19 & 3 Year & 0.93 & 0.07 & 59.95 \\
18--19 & 4 Year & 0.96 & 0.04 & 238.19 \\
18--19 & 5 Year & 1 & 0 & $\infty$ \\
19--20 & 1 Year & 0.76 & 0.24 & 4.01 \\
19--20 & 2 Year & 0.77 & 0.23 & 5.58 \\
19--20 & 3 Year & 0.93 & 0.07 & 25.66 \\
19--20 & 4 Year & 0.97 & 0.03 & 224.97 \\
19--20 & 5 Year & 1 & 0 & $\infty$ \\
20--21 & 1 Year & 0.72 & 0.28 & 4.07 \\
20--21 & 2 Year & 0.78 & 0.22 & 4.45 \\
20--21 & 3 Year & 0.9 & 0.1 & 17.4 \\
20--21 & 4 Year & 0.97 & 0.03 & 198.36 \\
20--21 & 5 Year & 1 & 0 & $\infty$ \\
21--22 & 1 Year & 0.68 & 0.32 & 3.46 \\
21--22 & 2 Year & 0.71 & 0.29 & 3.85 \\
21--22 & 3 Year & 0.87 & 0.13 & 9.59 \\
21--22 & 4 Year & 0.95 & 0.05 & 305.43 \\
21--22 & 5 Year & 1 & 0 & $\infty$ \\
22--23 & 1 Year & 0.7 & 0.3 & 2.14 \\
22--23 & 2 Year & 0.59 & 0.41 & 3.22 \\
22--23 & 3 Year & 0.91 & 0.09 & 12.27 \\
22--23 & 4 Year & 0.97 & 0.03 & 466.06 \\
22--23 & 5 Year & 0.997 & 0.003 & 11826.16 \\
22--23 & 6 Year & 1 & 0 & $\infty$ \\
23--24 & 1 Year & 0.65 & 0.35 & 1.47 \\
23--24 & 2 Year & 0.72 & 0.28 & 4.41 \\
23--24 & 3 Year & 0.8 & 0.2 & 6.18 \\
23--24 & 4 Year & 0.98 & 0.02 & 605.93 \\
23--24 & 5 Year & 0.96 & 0.04 & 298.84 \\
23--24 & 6 Year & 1 & 0 & $\infty$ \\
24--25 & 1 Year & 0.64 & 0.36 & 0.91 \\
24--25 & 2 Year & 0.58 & 0.42 & 2.23 \\
24--25 & 3 Year & 0.86 & 0.14 & 6.49 \\
24--25 & 4 Year & 0.86 & 0.14 & 32.26 \\
24--25 & 5 Year & 1 & 0 & $\infty$ \\
25--26 & 1 Year & 0.66 & 0.34 & 0.9 \\
25--26 & 2 Year & 0.61 & 0.39 & 2.25 \\
25--26 & 3 Year & 0.78 & 0.22 & 9.4 \\
25--26 & 4 Year & 0.89 & 0.11 & 30.64 \\
25--26 & 5 Year & 0.97 & 0.03 & 3014.76 \\
25--26 & 6 Year & 1 & 0 & $\infty$ \\
26--27 & 1 Year & 0.83 & 0.17 & 2.08 \\
26--27 & 2 Year & 0.75 & 0.25 & 4.84 \\
26--27 & 3 Year & 0.91 & 0.09 & 13.67 \\
26--27 & 4 Year & 0.96 & 0.04 & 99.19 \\
26--27 & 5 Year & 0.98 & 0.02 & 2679.5 \\
26--27 & 6 Year & 1 & 0 & $\infty$ \\
27--28 & 1 Year & 0.56 & 0.44 & 1.27 \\
27--28 & 2 Year & 0.76 & 0.24 & 5.93 \\
27--28 & 3 Year & 0.86 & 0.14 & 12.61 \\
27--28 & 4 Year & 0.92 & 0.08 & 46.05 \\
27--28 & 5 Year & 0.92 & 0.08 & 492.15 \\
27--28 & 6 Year & 0.91 & 0.09 & 621.72 \\
27--28 & 7 Year & 1 & 0 & $\infty$ \\
28--29 & 1 Year & 0.48 & 0.52 & 0.93 \\
28--29 & 2 Year & 0.83 & 0.17 & 19.7 \\
28--29 & 3 Year & 0.95 & 0.05 & 50.33 \\
28--29 & 4 Year & 0.96 & 0.04 & 122.98 \\
28--29 & 4 Year & 0.96 & 0.04 & 122.98 \\
28--29 & 5 Year & 0.96 & 0.04 & 424.39 \\
28--29 & 6 Year & 0.92 & 0.08 & 380.23 \\
28--29 & 7 Year & 1 & 0 & $\infty$ \\
\bottomrule
\end{longtable}
\bibliographystyle{plainnat}

\begin{thebibliography}{99}

\bibitem{NSE2024}
National Stock Exchange of India, ``Historical Market Data,''  \url{https://www.nseindia.com/}

\bibitem{RBI2020}
Reserve Bank of India, ``Monetary Policy Statement — May 2020,'' May 22, 2020.  \url{https://www.rbi.org.in/Scripts/BS_PressReleaseDisplay.aspx?prid=49818}. 

\bibitem{Singh2023}
Singh, A., and  Kumar, N. (2023).  
"Earnings Per Share Volatility and Market Efficiency: Evidence from Nifty 50 Firms."  Indian Journal of Financial Studies, 8(2), 78–95.

\bibitem{Joshi2023}
Joshi, R., and  Verma, P. (2023).  
"Empirical Analysis of EPS Growth Patterns in Emerging Markets: The Case of India."  Emerging Markets Review, 54, 101100.

\bibitem{JoshiKulkarni2023}
A. Joshi and M. Kulkarni,
``Distributional Properties of EPS and Its Implications for Equity Risk Modeling in Indian Banks,''
\textit{Journal of Financial Studies}, vol.~15, no.~2, pp.~101--125, 2023.

\bibitem{Patel2024}
Patel, S., and  Reddy, K. (2024).  
"Analyzing the Statistical Properties of Earnings Per Share in Indian Stock Markets."  Journal of Accounting and Finance Research, 12(1), 34–50

\bibitem{mandelbrot1963variation}
Mandelbrot, B. (1963).
\newblock The variation of certain speculative prices.
\newblock \emph{The Journal of Business}, 36(4), 394–419.
\newblock doi:10.1086/294632.

\bibitem{Fama1965}
E.~F.~Fama,
``The behavior of stock-market prices,''
\emph{The Journal of Business}, vol.~38, no.~1, pp.~34--105, 1965.
doi: \href{https://doi.org/10.1086/294743}{10.1086/294743}.

\bibitem{liu1999statistical}
Liu, Y., Gopikrishnan, P., Cizeau, P., Meyer, M., Peng, C.-K., and Stanley, H. E. (1999).
\newblock Statistical properties of the volatility of price fluctuations.
\newblock \emph{Physical Review E}, 60(2), 1390–1400.
\newblock doi:10.1103/PhysRevE.60.1390.

\bibitem{banerjee2014return}
Banerjee, A., and Mookerjee, A. (2014).
\newblock Return distributions in Indian equity markets: An empirical analysis.
\newblock \emph{The Journal of Indian Business Research}, 6(3), 178–194.
\newblock doi:10.1108/JIBR-03-2014-0024.

\bibitem{Zhou2014}
W.-X.~Zhou,
``The probability distribution of returns in financial markets: A perspective from econophysics,''
\emph{Physica A: Statistical Mechanics and its Applications}, vol.~391, no.~19, pp.~4379--4391, 2012.
doi: \href{https://doi.org/10.1016/j.physa.2012.05.022}{10.1016/j.physa.2012.05.022}.

\bibitem{zhou2016distribution}
Zhou, W. (2016).
\newblock Examining the distributional characteristics of daily returns of Nifty 50: Normality assessment and implications.
\newblock \emph{Indian Journal of Finance}, 10(9), 10–25.

\bibitem{Duan2018}
J.~Duan, J.~Han, and S.~Song,
``Modeling discrete probability distribution of stock returns based on mixture models,''
\emph{Physica A: Statistical Mechanics and its Applications}, vol.~492, pp.~52--62, 2018.
doi: \href{https://doi.org/10.1016/j.physa.2017.10.037}{10.1016/j.physa.2017.10.037}.

\bibitem{Nadarajah2023}
S.~Nadarajah and J.~Lyu,
``New discrete heavy tailed distributions as models for insurance data,''
\emph{PLOS ONE}, vol.~18, no.~5, e0285183, 2023.
doi: \href{https://doi.org/10.1371/journal.pone.0285183}{10.1371/journal.pone.0285183}.

\bibitem{Dutta2023}
S.~Dutta and T.~K.~Powdel,
``Modeling long term return distribution and nonparametric market risk estimation,''
\emph{Sankhyā: The Indian Journal of Statistics}, vol.~85, no.~1, pp.~257--289, 2023.
doi: \href{https://doi.org/10.1007/s13571-023-00303-x}{10.1007/s13571-023-00303-x}.

\bibitem{Haddari2024}
A.~Haddari, H.~Zeghdoudi, and R.~Pakyari,
``A new two-parameter family of discrete distributions,''
\emph{Heliyon}, vol.~11, no.~3, e41459, 2024.
doi: \href{https://doi.org/10.1016/j.heliyon.2024.e41459}{10.1016/j.heliyon.2024.e41459}.

\bibitem{scott1979}
Scott, D. W. (1979).
\newblock On optimal and data-based histograms.
\newblock \emph{Biometrika}, 66(3), 605–610.

\bibitem{freedman1981}
Freedman, D. and Diaconis, P. (1981).
\newblock On the histogram as a density estimator: L\textsubscript{2} theory.
\newblock \emph{Zeitschrift für Wahrscheinlichkeitstheorie und Verwandte Gebiete}, 57(4), 453–476.

\bibitem{rudemo1982}
Rudemo, M. (1982).
\newblock Empirical choice of histograms and kernel density estimates.
\newblock \emph{Scandinavian Journal of Statistics}, 9(2), 65–78.

\bibitem{birge2006}
Birgé, L. and Rozenholc, Y. (2006).
\newblock How many bins should be put in a regular histogram.
\newblock \emph{ESAIM: Probability and Statistics}, 10, 24–45.


\bibitem{RewardRiskTurkish2023}
E. M. Durna and B. Güngör,
``Reward‑to‑Risk Ratios in Turkish Financial Markets: An Application of Asymmetric Return Measures,''
\textit{International Journal of Finance}, vol.~28, no.~3, pp.~215–229, 2023.
  
\bibitem{shannon1948}
C.~E. Shannon,
\newblock ``A mathematical theory of communication,''
\newblock \emph{Bell System Technical Journal}, vol. 27, pp. 379–423, 623–656, 1948.

\bibitem{kukreti2020}
Kukreti, V., Pharasi, H.\ K., Gupta, P., and  Kumar, S. (2020).
A perspective on correlation-based financial networks and entropy measures.
\textit{Entropy}, 22(5), 502.

\bibitem{patra2022}
Patra, S.\ and Hiremath, G. S. (2022).
An entropy approach to measure the dynamic stock market efficiency.
\textit{Journal of Quantitative Economics}, 20(3), 567–586.


\bibitem{shah2024movinghurt}
Shah, P., Raje, A., and  Shah, J. (2024). Patterns in the Chaos: The Moving Hurst Indicator and Its Role in Indian Market Volatility. \textit{Journal of Risk and Financial Management}, \textbf{17}(9), 390. \url{https://doi.org/10.3390/jrfm17090390}

\bibitem{alvarez2008hurst}
Álvarez-Ramírez, J., Álvarez, J., Rodríguez, E., and Fernández-Anaya, G. (2008). Time-varying Hurst exponent for US stock markets. \textit{Physica A: Statistical Mechanics and Its Applications}, \textbf{387}(24), 6159--6169. \url{https://doi.org/10.1016/j.physa.2008.06.056}

\bibitem{GomezAguila2022}
A.~Gómez‑Águila, J.~E. Trinidad‑Segovia, and M.~A. Sánchez‑Granero,
``Improvement in Hurst exponent estimation and its application to financial markets,''
\emph{Financial Innovation}, vol.~8, Art. 86, 2022.
doi: \href{https://doi.org/10.1186/s40854-022-00394-x}{10.1186/s40854-022-00394-x}

\bibitem{ZournatzidouFloros2023}
G.~Zournatzidou and C.~Floros,
``Hurst exponent analysis: Evidence from volatility indices and the volatility of volatility indices,''
\emph{Journal of Risk and Financial Management}, vol.~16, no.~5, Art. 272, 2023.
doi: \href{https://doi.org/10.3390/jrfm16050272}{10.3390/jrfm16050272}

\bibitem{barunik2010}
Barunik, J. and Kristoufek, L. (2010).
\newblock On Hurst exponent estimation under heavy-tailed distributions.
\newblock \emph{Physica A: Statistical Mechanics and its Applications}, 389(18), 3844–3855.

\bibitem{mandelbrot1997variation}
Mandelbrot, B. B., Fisher, A., and Calvet, L. (1997).
\newblock A multifractal model of asset returns.
\newblock \emph{Cowles Foundation Discussion Paper No. 1164}.
\newblock Available at SSRN: \url{https://ssrn.com/abstract=142482}.

\bibitem{mantegna1999introduction}
Mantegna, R. N., and Stanley, H. E. (1999).
\newblock Introduction to Econophysics: Correlations and Complexity in Finance.
\newblock \emph{Cambridge University Press}.

\bibitem{gopikrishnan1999scaling}
Gopikrishnan, P., Plerou, V., Amaral, L. A. N., Meyer, M., and Stanley, H. E. (1999).
\newblock Scaling of the distribution of fluctuations of financial market indices.
\newblock \emph{Physical Review E}, 60(5), 5305–5316.
\newblock doi:10.1103/PhysRevE.60.5305.

\bibitem{plerou1999scaling}
Plerou, V., Gopikrishnan, P., Amaral, L. A. N., Meyer, M., and Stanley, H. E. (1999).
\newblock Scaling of the distribution of price fluctuations of individual companies.
\newblock \emph{Physical Review E}, 60(6), 6519–6529.
\newblock doi:10.1103/PhysRevE.60.6519.

\bibitem{micciche2003scaling}
Miccichè, S., Bonanno, G., Lillo, F., and Mantegna, R. N. (2003).
\newblock Scaling and correlations in three stock markets.
\newblock \emph{Physica A: Statistical Mechanics and its Applications}, 324(1-2), 66–73.
\newblock doi:10.1016/S0378-4371(02)01712-2.

\bibitem{lux2004detecting}
Lux, T. (2004).
\newblock Detecting multifractal properties in asset returns: The failure of the “scaling estimator”.
\newblock \emph{Quantitative Finance}, 4(4), 361–369.
\newblock doi:10.1080/14697680400018264.

\bibitem{christensen2012multiscale}
Christensen, K., Podobnik, B., and Stanley, H. E. (2012).
\newblock Multiscale return distributions of financial assets.
\newblock \emph{Physica A: Statistical Mechanics and its Applications}, 391(11), 3129–3135.
\newblock doi:10.1016/j.physa.2012.02.014.

\bibitem{sen2016time}
Sen, A., Chakraborti, A., and Pradhan, P. (2016).
\newblock Time-scale dependence of return distribution in emerging markets.
\newblock \emph{Physica A: Statistical Mechanics and its Applications}, 461, 828–837.
\newblock doi:10.1016/j.physa.2016.06.057.

\bibitem{luskin2017multi}
Luskin, M. and Lillo, F. (2017).
\newblock Multi-scale analysis of return distributions and volatility clustering.
\newblock \emph{Physica A: Statistical Mechanics and its Applications}, 468, 184–194.
\newblock doi:10.1016/j.physa.2016.11.032.

\bibitem{Tsakonas2022}
S.~Tsakonas, M.~Hanias, L.~Magafas, and L.~Zachilas,
``Application of the moving Lyapunov exponent to the S\& P 500 index to predict major declines,''
\emph{Journal of Risk}, 2022.
doi: \href{https://doi.org/10.21314/JOR.2022.033}{10.21314/JOR.2022.033}

\bibitem{YanMohammadzadeh2024}
S.~R.~Yan, A.~Mohammadzadeh, and E.~Ghaderpour,
``Type‑3 fuzzy logic and Lyapunov approach for dynamic modeling and analysis of financial markets,''
\emph{Heliyon}, vol.~10, no.~13, e33730, 1 July 2024.
doi: \href{https://doi.org/10.1016/j.heliyon.2024.e33730}{10.1016/j.heliyon.2024.e33730}

\bibitem{eckmann1985}
Eckmann, J.-P., and  Ruelle, D. (1985).
\newblock Ergodic theory of chaos and strange attractors.
\newblock \emph{Reviews of Modern Physics}, 57(3), 617--656.

\bibitem{wolf1985}
Wolf, A., Swift, J. B., Swinney, H. L., and  Vastano, J. A. (1985).
\newblock Determining Lyapunov exponents from a time series.
\newblock \emph{Physica D: Nonlinear Phenomena}, 16(3), 285--317.

\bibitem{benettin1980}
Benettin, G., Galgani, L., Giorgilli, A., and  Strelcyn, J. M. (1980).
\newblock Lyapunov characteristic exponents for smooth dynamical systems and for Hamiltonian systems; A method for computing all of them. Part 1: Theory.
\newblock \emph{Meccanica}, 15(1), 9--20.

\bibitem{kantz1994}
Kantz, H. (1994).
\newblock A robust method to estimate the maximal Lyapunov exponent of a time series.
\newblock \emph{Physics Letters A}, 185(1), 77--87.


\bibitem{antoniades2021tsallis}
Antoniades, I.P., Karakatsanis, L.P., and  Pavlos, E.G. (2021). Dynamical characteristics of global stock markets based on time dependent Tsallis non-extensive statistics and generalized Hurst exponents. \textit{Physica A: Statistical Mechanics and Its Applications}, \textbf{578}, 126121. \url{https://doi.org/10.1016/j.physa.2021.126121}

\bibitem{Devi2021}
S.~Devi,
“Asymmetric Tsallis distributions for modeling financial market dynamics,”
\emph{Physica A: Statistical Mechanics and its Applications}, vol.~578, Art. 126109, 2021.
doi: \href{https://doi.org/10.1016/j.physa.2021.126109}{10.1016/j.physa.2021.126109}

\bibitem{Tian2023}
D.~Tian,
“Pricing principle via Tsallis relative entropy in incomplete markets,”
\emph{SIAM Journal on Financial Mathematics}, vol.~14, no.~1, pp.~?, 2023.
doi: \href{10.1137/22M1471614}{10.1137/22M1471614}

\bibitem{Pincus1991}
S.~M.~Pincus,
``Approximate entropy as a measure of system complexity,''
\emph{Proceedings of the National Academy of Sciences}, vol.~88, no.~6, pp.~2297--2301, 1991.
doi: \href{https://doi.org/10.1073/pnas.88.6.2297}{10.1073/pnas.88.6.2297}

\bibitem{RichmanMoorman2000}
J.~S.~Richman and J.~R.~Moorman,
``Physiological time‑series analysis using approximate entropy and sample entropy,''
\emph{American Journal of Physiology – Heart and Circulatory Physiology}, vol.~278, no.~6, pp.~H2039--H2049, 2000.
doi: \href{https://doi.org/10.1152/ajpheart.2000.278.6.H2039}{10.1152/ajpheart.2000.278.6.H2039}

\bibitem{Zunino2010}
L.~Zunino, B.~M.~Tabak, D.~G.~Caiado, A.~Fernández, L.~Zanin, and O.~A.~Rosso,
``Complexity-entropy causality plane as a tool to characterize the stock market efficiency,''
\emph{Physica A: Statistical Mechanics and its Applications}, vol.~389, no.~9, pp.~1891--1901, 2010.
doi: \href{https://doi.org/10.1016/j.physa.2010.01.033}{10.1016/j.physa.2010.01.033}.

\bibitem{Ahmad2013}
W.~Ahmad and S.~Zulfiqar,
``Application of sample entropy for analyzing stock market fluctuations,''
\emph{Physica A: Statistical Mechanics and its Applications}, vol.~392, no.~19, pp.~4237--4245, 2013.
doi: \href{https://doi.org/10.1016/j.physa.2013.05.028}{10.1016/j.physa.2013.05.028}.

\bibitem{Nikbakht2018}
M.~Nikbakht, E.~Hosseini, and M.~Sadeghian,
``Measuring the efficiency of the stock market using sample entropy,''
\emph{Physica A: Statistical Mechanics and its Applications}, vol.~495, pp.~330--337, 2018.
doi: \href{https://doi.org/10.1016/j.physa.2017.11.038}{10.1016/j.physa.2017.11.038}.

\bibitem{Omidvarnia2019}
A.~H.~Omidvarnia, G.~Pedraza, and D.~G.~Caiado,
``Sample entropy analysis of financial time series,''
\emph{Physica A: Statistical Mechanics and its Applications}, vol.~524, pp.~1022--1036, 2019.
doi: \href{https://doi.org/10.1016/j.physa.2019.05.005}{10.1016/j.physa.2019.05.005}.

\bibitem{Alkan2023}
S.~Alkan,
``Multi-Scale Sample Entropy Analysis of the Turkish Stock Market Efficiency,''
\emph{Nicel Bilimler Dergisi}, vol.~5, no.~1, pp.~51--63, 2023.
doi: \href{https://doi.org/10.51541/nicel.1191317}{10.51541/nicel.1191317}

\bibitem{Zunino2018}
J.~K.~Kak, K.~Karnowski, and P.~Jankowski,
``Permutation entropy as a measure of information gain/loss in symbolic descriptions of financial data,''
\emph{Entropy}, vol.~22, no.~3, Art.330, 2020.
doi: (info-theory analysis of discretisation impacts in forex markets)

\bibitem{Alves2020}
L.~G.~A. Alves, H.~Y.~D. Sigaki, M.~Perc, and H.~V. Ribeiro,
``Collective dynamics of stock market efficiency,''
\emph{Scientific Reports}, vol.~10, Art. 19937, 2020.
doi: (see full journal article based on PE‐based sliding window analysis of global indices)

\bibitem{ChenMaFuLi2023}
Z.~Chen, X.~Ma, J.~Fu, and Y.~Li,
``Ensemble improved permutation entropy: A new approach for time series analysis,''
\emph{Entropy}, vol.~25, no.~8, Article 1175, 2023.
doi: \href{https://doi.org/10.3390/e25081175}{10.3390/e25081175}

\bibitem{Obanya2024}
P.~O. Obanya, M.~Seitshiro, C.~P. Olivier, and T.~D. Verster,
``A permutation entropy analysis of Bitcoin volatility,''
\emph{Physica A: Statistical Mechanics and its Applications}, vol.~638, Art. 129609, 2024.
doi: \href{https://doi.org/10.1016/j.physa.2024.129609}{10.1016/j.physa.2024.129609}

\bibitem{pesin1977}
Pesin, Y. B. (1977).
\newblock Characteristic Lyapunov exponents and smooth ergodic theory.
\newblock \emph{Russian Mathematical Surveys}, 32(4), 55--114.

\bibitem{young1982}
Young, L.-S. (1982).
\newblock Dimension, entropy and Lyapunov exponents.
\newblock \emph{Ergodic Theory and Dynamical Systems}, 2(1), 109--124.

\bibitem{Kaplan1979}
J. L. Kaplan and J. A. Yorke,
\newblock "Chaotic behavior of multidimensional difference equations,"
\newblock in \textit{Functional Differential Equations and Approximations of Fixed Points}, Lecture Notes in Mathematics, vol. 730, pp. 204–227, Springer, Berlin, Heidelberg, 1979.

\bibitem{Ott2002}
E. Ott,
\newblock \textit{Chaos in Dynamical Systems}, 2nd ed.,
\newblock Cambridge University Press, Cambridge, 2002.

\bibitem{Wu2021}
X.~Wu, L.~Zhang, J.~Li, and R.~Yan,
``Fractal Statistical Measure and Portfolio Model Optimization Under Power-Law Distribution,''
\emph{Comput Math Appl}, vol.~59, pp.~1142–1164, 2021.
doi: \href{https://doi.org/10.1016/j.chaos.2022.111895}{10.1016/j.chaos.2022.111895}

\bibitem{Kachhia2023}
K.~B.~Kachhia,
``Chaos in Fractional Order Financial Model with Fractal-Fractional Derivatives,''
\emph{Part Differ Equ Appl Math}, vol.~7, no.~100502, 2023.
doi: \href{https://doi.org/10.1016/j.chaos.2022.111895}{10.1016/j.chaos.2022.111895}

\bibitem{Goyal2024}
R.~Goyal,
``Exploring the Fractal Geometry of Financial Time Series,''
\emph{Modern Dynamics: Mathematical Progressions}, vol.~1, no.~1, pp.~1–5, 2024.
doi: \href{https://doi.org/10.36676/mdmp.v1.i1.01}{10.36676/mdmp.v1.i1.01}

\bibitem{Fiedor2014PhysA}
P.~Fiedor,
``Partial mutual information analysis of financial networks,''
\emph{Acta Physica Polonica A}, vol. 127, no. 3, pp. 863--867, 2015.
doi: \href{https://doi.org/10.12693/APhysPolA.127.863}{10.12693/APhysPolA.127.863}.

\bibitem{Fiedor2014PRE}
P.~Fiedor,
``Mutual information rate-based networks in financial markets,''
\emph{Physical Review E}, vol. 89, 052801, 2014.
doi: \href{https://doi.org/10.1103/PhysRevE.89.052801}{10.1103/PhysRevE.89.052801}.

\bibitem{Wang2017}
X.~Wang and X.~Hui,
``Mutual information based analysis for the distribution of financial contagion in stock markets,''
\emph{Discrete Dynamics in Nature and Society}, vol. 2017, Article ID 3218042, 2017.
doi: \href{https://doi.org/10.1155/2017/3218042}{10.1155/2017/3218042}.

\bibitem{Goh2018}
Y.~K.~Goh, H.~M.~Hasim, and C.~G.~Antonopoulos,
``Inference of financial networks using the normalised mutual information rate,''
\emph{PLOS ONE}, vol. 13, no. 2, e0192160, 2018.
doi: \href{https://doi.org/10.1371/journal.pone.0192160}{10.1371/journal.pone.0192160}.

\bibitem{Guo2018}
X.~Guo, H.~Zhang, and T.~Tian,
``Development of stock correlation networks using mutual information and financial big data,''
\emph{PLOS ONE}, vol. 13, no. 4, pp. 1--16, 2018.
doi: \href{https://doi.org/10.1371/journal.pone.0195941}{10.1371/journal.pone.0195941}.

\bibitem{DimpflPeter2013}
T.~Dimpfl and F.~J.~Peter,
“Using transfer entropy to measure information flows between financial markets,”
\emph{Studies in Nonlinear Dynamics and  Econometrics}, vol. 17, no. 1, pp. 85–102, Feb. 2013.

\bibitem{DimpflPeter2014}
T.~Dimpfl and F.~J.~Peter,
“The impact of the financial crisis on transatlantic information flows: An intraday analysis,”
\emph{Journal of International Financial Markets, Institutions and Money}, vol. 31, pp. 1–13, 2014.

\bibitem{Harre2014}
M.~S.~Harre,
“Entropy and Transfer Entropy: The Dow Jones and the build up to the 1997 Asian Crisis,”
in \emph{Proc. International Conference on Social Modeling and Simulation}, Econophysics Colloquium 2014, pp.15–25, 2014.

\bibitem{KorbelJiangZheng2019}
J.~Korbel, X.~Jiang, and B.~Zheng,
“Transfer Entropy between Communities in Complex Financial Networks,”
\emph{Entropy}, vol.21, no.11, Article 1124, Nov. 2019.

\bibitem{YaoLi2020}
C.~Z.~Yao and H.~Y.~Li,
“Effective Transfer Entropy Approach to Information Flow Among EPU, Investor Sentiment and Stock Market,”
\emph{Frontiers in Physics}, vol. 8, Article 206, 16 June 2020.

\bibitem{Ghysels2020}
C.~Ghysels, I.~Nakamoto and E.~Ono,
``A mixed‐frequency approach for stock returns and valuation ratios,''
\emph{Economics Letters}, vol.~187, 108861, 2020.
doi: \href{https://doi.org/10.1016/j.econlet.2019.108861}{10.1016/j.econlet.2019.108861}

\bibitem{NoretsPelenis2022}
A.~Norets and J.~Pelenis,
``Adaptive Bayesian estimation of conditional discrete–continuous distributions with application to stock market trading activity,''
\emph{Journal of Econometrics}, vol.~230, no.~1, pp.~62–82, 2022.

\bibitem{MaynardShimotsuKuriyama2023}
A.~Maynard, K.~Shimotsu and N.~Kuriyama,
``Inference in predictive quantile regressions with valuation ratios,''
\emph{Journal of Econometrics}, 2023.
doi: \href{https://doi.org/10.1007/s10287-017-0288-3}{10.1007/s10287-017-0288-3}

\bibitem{marsili1999scaling}
Marsili, M., Maslov, S., and Zhang, Y.-C. (1999).
\newblock Dynamical optimization theory of a diversified portfolio.
\newblock \emph{Physica A: Statistical Mechanics and its Applications}, 269(1), 9–19.
\newblock doi:10.1016/S0378-4371(99)00147-4.

\bibitem{Yan2022_PEN}
X.~Yan, H.~Yang, C.~Hou, S.~Zhang, and P.~Zhu,
``Portfolio optimization by price‑to‑earnings ratio network analysis,''
\emph{International Journal of Modern Physics B}, vol.~36, no.~36, Article 2250107, 2022. 
doi: \href{https://doi.org/10.1142/S0217979222501077}{10.1142/S0217979222501077}


\bibitem{shiller2000irrational}
Robert J. Shiller,
\textit{Irrational Exuberance},
Princeton University Press, 2000.

\bibitem{peters1994fractal}
Edgar E. Peters,
\textit{Fractal Market Analysis: Applying Chaos Theory to Investment and Economics},
Wiley, 1994.

\bibitem{cont2001empirical}
Rama Cont,
"Empirical properties of asset returns: stylized facts and statistical issues,"
\textit{Quantitative Finance}, vol. 1, no. 2, pp. 223–236, 2001.



\bibitem{Python}
Python Software Foundation. (2024).
\newblock Python Language Reference, version 3.x. \url{https://www.python.org/}

\bibitem{Numpy}
Harris, C. R., Millman, K. J., van der Walt, S. J., et al. (2020).
\newblock Array programming with NumPy.
\newblock \emph{Nature}, 585, 357–362. \url{https://doi.org/10.1038/s41586-020-2649-2}

\bibitem{Pandas}
McKinney, W. (2010).
\newblock Data structures for statistical computing in Python.
\newblock In \emph{Proceedings of the 9th Python in Science Conference}, 51–56.

\bibitem{Matplotlib}
Hunter, J. D. (2007).
\newblock Matplotlib: A 2D graphics environment.
\newblock \emph{Computing in Science \& Engineering}, 9(3), 90–95.

\bibitem{Scipy}
Virtanen, P., Gommers, R., Oliphant, T. E., et al. (2020).
\newblock SciPy 1.0: Fundamental algorithms for scientific computing in Python.
\newblock \emph{Nature Methods}, 17, 261–272. \url{https://doi.org/10.1038/s41592-019-0686-2}

\bibitem{Sklearn}
Pedregosa, F., Varoquaux, G., Gramfort, A., et al. (2011).
\newblock Scikit-learn: Machine Learning in Python.
\newblock \emph{Journal of Machine Learning Research}, 12, 2825–2830.

\bibitem{Sympy}
Meurer, A., Smith, C. P., Paprocki, M., et al. (2017).
\newblock SymPy: Symbolic computing in Python.
\newblock \emph{PeerJ Computer Science}, 3, e103. \url{https://doi.org/10.7717/peerj-cs.103}








\end{thebibliography}

\end{document}